\documentclass[journal]{IEEEtran}
\usepackage{latexsym}
\usepackage{amsfonts,amssymb,amsmath}
\usepackage{hyperref}
\def\BibTeX{{\rm B\kern-.05em{\sc i\kern-.025em b}\kern-.08em T\kern-.1667em\lower.7ex\hbox{E}\kern-.125emX}}
\usepackage{algorithm}
\usepackage{algpseudocode}
\usepackage{amsmath}
\usepackage{amsfonts}
\usepackage{graphicx}
\usepackage{epsfig}
\usepackage{cite}
\usepackage{float}
\usepackage{txfonts}
\usepackage{relsize}
\usepackage{color}
\usepackage{color}
\usepackage{CJK}
\usepackage{times}
\usepackage{multicol}
\usepackage{stfloats}
\usepackage{float}
\usepackage{lipsum}
\usepackage{bm}

\begin{document}

\title{Fluid Antenna System-Enabled Mitigation of Asynchronous Reception in Cell-Free Massive MIMO Systems
\thanks{This work was supported by the Hong Kong Research Grants Council with Area of Excellence grant AoE/E-601/22-R.}}
\author{Jun~Qian,~\IEEEmembership{Member,~IEEE,}~Zan~Li,~\IEEEmembership{Member,~IEEE,}~Junhui~Rao,~\IEEEmembership{Member,~IEEE,}
Ross~Murch,~\IEEEmembership{Fellow,~IEEE,}~and~Khaled~B.~Letaief,~\IEEEmembership{Fellow,~IEEE}

\thanks{The authors are with the Department of Electronic and Computer Engineering, The Hong Kong University of Science and Technology,
Hong Kong (e-mail: eejunqian@ust.hk, zligq@connect.ust.hk, jraoaa@connect.ust.hk, eermurch@ust.hk, eekhaled@ust.hk).}}

\maketitle
{\begin{abstract}

Distributed cell-free massive multiple-input multiple-output (MIMO) systems offer immense potential for sixth-generation (6G) networks. However, practical distributed deployments inherently suffer from asynchronous signal arrivals, which exacerbate multi-user interference and degrade system performance, especially for coherent transmission. To natively mitigate the asynchronous reception effect, this paper proposes integrating fluid antenna systems (FASs) into distributed cell-free massive MIMO systems, exploiting their reconfigurable spatial positions to release additional spatial degrees of freedom (DoFs). We establish the FAS-enabled data transmission model with asynchronous reception, i.e., delay phases. We also derive the analytical downlink spectral efficiency (SE) performance of the proposed system under coherent and non-coherent transmissions, using low-complexity Maximum Ratio (MR) precoding to provide fundamental theoretical bounds. Specifically, we propose a novel nonmonotone accelerated projected gradient ascent algorithm to jointly optimize FAS positions and power control coefficients, maximizing the downlink sum SE. Numerical results demonstrate that while asynchronous reception severely degrades system performance for coherent transmission, the spatial DoFs unlocked by optimized FAS positions, along with efficient power control, can significantly counteract the effects of unknown delay phases and outperform traditional fixed-position antennas. For non-coherent transmission, which inherently bypasses asynchronous reception, the application of FAS leverages spatial reconfigurability to natively maximize signal strength and achieve more pronounced SE gains. Ultimately, our proposed FAS-enabled system, coupled with efficient power control, mitigates performance degradation due to asynchronous reception and outperforms traditional fixed-position antennas, paving the way for the practical deployment of FASs in robust, highly efficient 6G cell-free massive MIMO systems.

\end{abstract}

}

\maketitle

\section{Introduction}

\IEEEPARstart{T}{he} ever-increasing demand for ubiquitous connectivity and higher data rates for future sixth-generation (6G) wireless communication has positioned cell-free massive multiple-input multiple-output (MIMO) as a promising technology, with the potential to provide more extensive coverage, favorable propagation configurations and diminished inter-cell interference \cite{7827017,10201892,10422885,9665300}. By evolving from traditional cellular structures to a distributed architecture, cell-free massive MIMO coordinates numerous geographically dispersed access points (APs) via a central processing unit (CPU) to jointly serve users within a wide area\cite{7917284,9416909,11206383,8676341,11196010}. Based on the law of large numbers, favorable propagation and channel hardening can characterize the propagation environment of cell-free massive MIMO systems, facilitating simple precoding design and interference management \cite{10032129}.
With the topology of integrating distributed networks and massive MIMO, cell-free massive MIMO systems can provide rich macro-diversity gains to greatly improve spectral efficiency (SE) performance\cite{10201892,7827017,10297571,8676341}. The authors of \cite{7827017} indicated that cell-free massive MIMO systems could outperform small-cell systems by achieving a higher $95\%$ SE per user, thanks to joint interference cancellation. Furthermore, joint signal processing in cell-free massive MIMO systems can effectively mitigate the impact of non-ideal practical factors by introducing additional spatial degrees of freedom (DoFs), thereby enabling spatial multiplexing gains, capitalizing on the advantages of cell-free massive MIMO and proving to be practical and scalable\cite{10201892,10032129,9875036,10167480}. 

Typically, coherent transmission in cell-free massive MIMO systems benefits from distributing transmit power across multiple APs, even when APs exhibit different channel gains for users \cite{8809413}. However, realizing the full potential of coherent joint transmission fundamentally relies on the assumption of perfect time synchronization across all distributed APs\cite{10201892,7827017,11196010,11206383}. This assumption is challenging for practical distributed networks \cite{10319677,8676341,9531352,10032129}. The geographically distributed nature of the APs results in varying propagation distances for each user, leading to inevitable time differences in signal arrival times \cite{10032129,8676341}. Consequently, asynchronous
reception effects in terms of delay phases will occur during signal demodulation \cite{10032129,8676341,10319677}. The incurred delay phases degrade the quality of the received signals and pose challenges to achieving coherent transmission in cell-free massive MIMO systems \cite{9531352,10319677}. To tackle these limitations, \cite{4287203}
addressed the asynchronous signal reception by extending the cyclic prefix (CP) duration, at the cost of per-user net throughput. \cite{10032129} implemented rate splitting to alleviate performance degradation from delay phases and oscillator phases on cell-free massive MIMO systems. \cite{10319677,8676341} explored the effects of asynchronous reception in distributed
massive MIMO orthogonal frequency-division multiplexing (OFDM) systems. \cite{10804196} proposed a mixed coherent and non-coherent transmission scheme to reduce the effects of asynchronous reception in cell-free MIMO-OFDM systems. Despite these efforts, developing sophisticated technologies remains crucial to addressing asynchronous reception challenges in robust cell-free massive MIMO systems and achieving ubiquitous connectivity. Fortunately, thanks to the development of advanced antenna technologies, fluid antenna systems (FASs) \cite {10318134,10146262}, which offer the flexibility of dynamic antenna movement and additional spatial DoFs, can be considered a potential approach to compensate for the inherent asynchronous reception caused by the distributed nature of APs in cell-free massive MIMO systems.

The FAS, also known as movable antennas, has been introduced as groundbreaking technology for 6G wireless networks to enhance diversity and spatial multiplexing gains \cite{10318134,10146262,10146286,10146274,10839251}. FAS utilizes a software-controllable
fluidic, conductive, or dielectric structure comprising multiple preset positions (known as ports) and one radio frequency (RF) chain. By dynamically adjusting the antenna position within a designated local region, FAS can physically reconfigure the effective channel, thereby unlocking additional spatial DoFs to maximize diversity and multiplexing gains\cite{10328751,10318134,9650760}. Subsequently,
\cite{9650760} proposed fluid antenna multiple access for multi-user communications, achieving promising rates and reliability. \cite{10066316} introduced slow fluid antenna multiple access (s-FAMA), updating FAS positions only if the fading
channel changes to go beyond the impractical symbol-by-symbol basis. \cite{10103838} refined the channel model to accurately approximate the FAS-enabled channel distribution.  
 \cite{10318134} explored the applicability of FAS in both orthogonal multiple access and non-orthogonal multiple access networks, focusing on optimal position selection and power allocation to improve performance. The authors of \cite{10146262} adopted reconfigurable intelligent surfaces (RISs) to create a rich scattering
environment, enabling FAS to mitigate multi-user interference. \cite{11016053} studied FAS-assisted integrated
sensing and communication (ISAC) systems, optimizing antenna position and beamforming to maximize the sensing signal-to-interference-and-noise-ratio (SINR) with varying channel state information (CSI). Similarly, \cite{10839251} leveraged movable antennas in ISAC to maximize the communication rate and the sensing mutual information via beamforming optimization.

In light of the flexibility and practicality of FAS, recent works have begun exploring the performance of FAS incorporated with other promising applications, including the immense potential to enhance cell-free massive MIMO systems. \cite{10694457} studied the uplink power
control and antenna position optimization in user-centric cell-free massive MIMO systems with FAS-enabled APs. \cite{10827177}
examined cell-free symbiotic radio systems featuring movable antenna-enabled APs, highlighting the benefits of this integration in the presence of an eavesdropper. 
\cite{10891142} proposed a six-dimensional movable
antenna-aided cell-free massive MIMO system to fully exploit both micro and macro spatial diversities. \cite{11018493} introduced a dynamic neighborhood pruning particle swarm optimization algorithm to optimize transmit beamforming and movable antenna positions jointly in movable
antenna-aided cell-free massive MIMO systems. \cite{11351331} studied the FAS-enabled wireless power transfer in cell-free massive MIMO systems. \cite{11353414} considered the hardware-related time synchronization errors at transmitters in the cell-free OFDM system and utilized FAS-enabled APs to mitigate the errors. However, the investigation of FAS within cell-free massive MIMO systems still remains nascent; for example, the inherent asynchronous reception caused by the distributed nature of APs and users has not been analyzed. Therefore, these observations motivate our work to analyze the performance of cell-free massive MIMO systems with asynchronous reception due to physical distances, and to explore the potential of leveraging the spatial flexibility of FAS-enabled users to counteract the destructive effects of asynchronous reception.

Motivated by these observations, this paper investigates the downlink SE performance of cell-free massive MIMO systems where users are equipped with FAS and experience asynchronous reception. We develop a comprehensive downlink data transmission model with asynchronous reception, characterized by delay phases, encompassing both coherent and non-coherent transmission. We implement maximum-ratio (MR) precoding to establish a fundamental theoretical baseline for the spatial gains enabled by FAS. Furthermore, we propose a novel nonmonotone accelerated projected gradient ascent (NAPGA) algorithm to jointly optimize the FAS positions and power control coefficients, maximizing the downlink sum SE. Then, we summarize the key contributions of this work as follows.

\begin{itemize}
    \item We establish a downlink channel model for the FAS-enabled cell-free massive MIMO system incorporating asynchronous reception characterized by delay phases. This contrasts with traditional fixed-position antennas by introducing dynamic spatial flexibility through adjustable antenna positions at FAS-enabled users.

    \item To investigate the effects of asynchronous reception, we derive the analytical downlink SE expressions for both coherent and non-coherent downlink transmission. By incorporating low-complexity MR precoding, we systematically quantify the severe performance degradation induced by delay phases in coherent transmission, while establishing non-coherent transmission as an inherently delay phase-immune baseline to evaluate the comparative spatial gains unlocked by FAS.

    \item To fully exploit the spatial flexibility of FAS and offset the performance degradation caused by asynchronous reception, we establish a joint optimization problem for FAS positions alongside downlink power control coefficients. We solve this non-convex maximization problem utilizing a novel tailored NAPGA algorithm to help ameliorate the effect of asynchronous reception, referring to \cite{10387264,NIPS2015_f7664060,hu2025uplinktransmissiondesignfluid}.

    \item We numerically validate the proposed system performance and establish concrete system design guidelines. While asynchronous reception severely degrades traditional system capacity, our results demonstrate that optimizing FAS positions alongside efficient power control can alleviate these adverse effects, significantly outperforming traditional fixed-position antenna systems.

\end{itemize}
The remainder of this paper is structured as follows. Section II develops the cell-free massive MIMO channel model with FAS-enabled users and asynchronous reception. Section III details the downlink data transmission model for both coherent and non-coherent transmission. Section IV formulates the downlink sum SE maximization problem pertaining to FAS position optimization and power control optimization. Section V presents numerical results and discussion, and Section VI concludes the paper and outlines future work.

{\textit{Notation:} In this paper, ${\textbf H}^T$, ${\textbf H}^H$, ${\textbf H}^*$ and ${\textbf H}^{-1}$ represent the transpose, conjugate-transpose, conjugate and inverse of a matrix $\textbf H$, respectively. $|\cdot|$ and $||\cdot||$ denote the respective absolute value and standard norm. $ \mathcal{CN}\left( 0,\sigma^2 \right)$ is the circularly symmetric complex Gaussian distribution with zero mean and covariance $\sigma^2$. $\mathbb  E\{\cdot \}$ is the expectation operator, $\mathfrak{RE}\{\cdot \}$ is the real part, and $[\textbf{x}]_{+}$ is the projector onto
the positive orthant. }

\section{System Model}
\begin{figure}[t!]
    \centering
    \includegraphics[width=\linewidth]{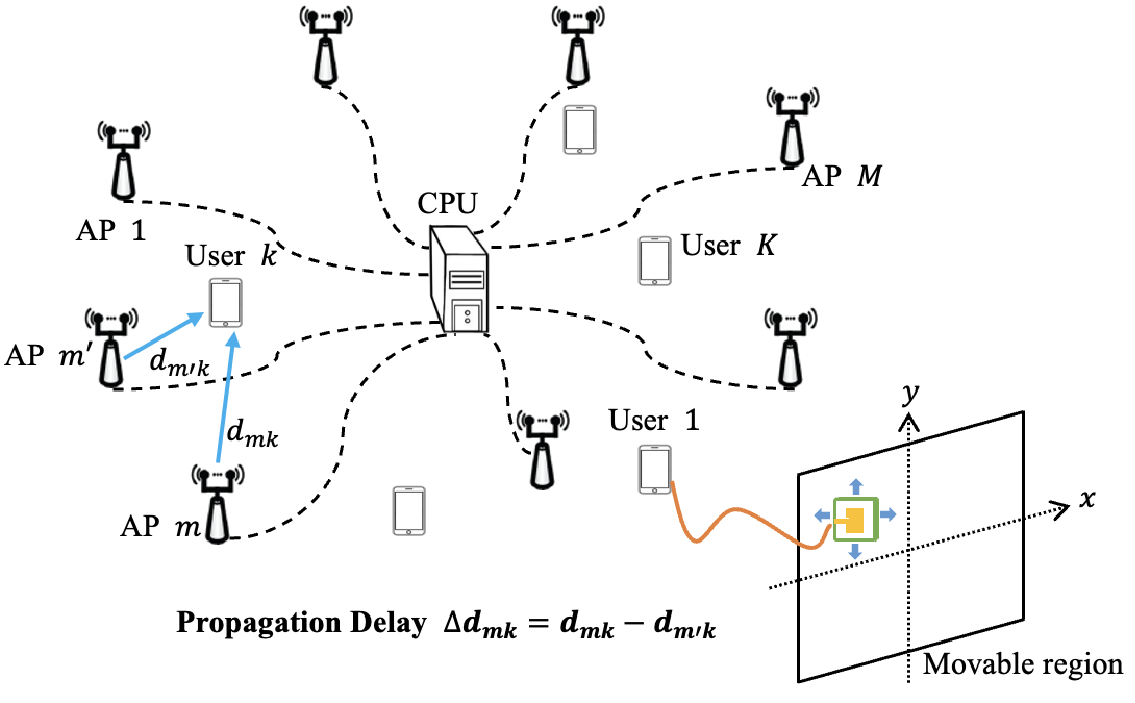}
    \caption{A diagram of the cell-free massive MIMO system with asynchronous reception and FAS-enabled users.}
    \label{system_model}
\end{figure}
\subsection{FAS-enabled Channel Model}

In this work, we consider a cell-free massive MIMO system operating in time-division duplex (TDD) mode, comprising $M$ APs and $K$ single-antenna users randomly distributed across a geographic area \cite{11196010,11206383}, with all APs linked to the CPU via ideal backhaul links. Each AP is equipped with
a uniform planar array possessing $N=N_h \times N_v$ antennas, where
$N_h$ and $N_v$ denote the number of antennas along the horizontal
and vertical directions, respectively\cite{10827177}. Moreover, each user is equipped with a single two-dimensional (2D) surface FAS\cite{10827177,10328751}, 
connected to the
RF chain via a flexible cable to be freely adjusted within the local region $\mathcal{C}_k=[x_\text{min},x_\text{max}]\times[y_\text{min},y_\text{max}],~\forall k$
\cite{10328751}. In this work, we assume that 
a square region is adopted for FAS with $x_\text{min}=y_\text{min}=d_\text{min}$ and $x_\text{max}=y_\text{max}=d_\text{max}$.
The positions of the $n$-th antenna at the $m$-th AP and the FAS at the $k$-th user are respectively denoted as $\textbf{t}_{mn}=[x_{mn},y_{mn}]^T$ and $\textbf{u}_{k}=[x_{k},y_{k}]^T$, $\forall m,~\forall n,~\forall k$\cite{11018493}, following the 2D Cartesian coordinate system.
The reference points of the $k$-th user and the $m$-th AP are $O_k$ and $D_m$, respectively. 
The antenna position collection at the 
$m$-th AP is denoted
by $\textbf{t}_m=[\textbf{t}_{m1},\textbf{t}_{m2},...,\textbf{t}_{mN}]\in\mathbb{R}^{2\times N},~\forall m$, and the position collection of FAS is $\textbf{u}=[\textbf{u}_1,\textbf{u}_2,...,\textbf{u}_K]\in\mathbb{R}^{2\times K}$.
A far-field plane-wave-based model is used to model the field response, since the AP-user distance is much larger than the size of the FAS region and the AP aperture\cite{11018493}. The near-field spherical-wave-based channel model will be studied in the future. 
Accordingly, the signal propagation distance difference for the $l$-th path between the $n$-th antenna of the $m$-th AP and the $k$-th user is
$\rho_{k,mn}^l(\textbf{t}_{mn})=x_{mn}\text{sin}\theta^t_{mk,l}\text{cos}\phi^t_{mk,l}+y_{mn}\text{cos}\theta^t_{mk,l}$ \cite{10328751,10243545}.
Then, the signal propagation distance difference for the $l$-th receive channel path between the FAS position $\textbf{u}_{k}$, and the origin of the local coordinate system at the $k$-th user is given by
$\rho_{mk}^l(\textbf{u}_{k})=x_{k}\text{sin}\theta^r_{mk,l}\text{cos}\phi^r_{mk,l}+y_{k}\text{cos}\theta^r_{mk,l}$ \cite{10243545,10328751}.
The transmit field response vector between the $n$-th antenna at the $m$-th AP and the $k$-th user, $\textbf{g}_{mk,n}(\textbf{t}_{mn})\in\mathbb{C}^{L_{mk,t}\times1}$ is
\begin{equation}
   \textbf{g}_{mk,n}(\textbf{t}_{mn})=[e^{j\frac{2\pi}{\lambda}\rho_{k,mn}^1(\textbf{t}_{mn})},e^{j\frac{2\pi}{\lambda}\rho_{k,mn}^2(\textbf{t}_{mn})},...,e^{j\frac{2\pi}{\lambda}\rho_{k,mn}^{L_{mk,t}}(\textbf{t}_{mn})}]^T,
\end{equation}
 and the transmit field response vector between the $k$-th user and the $m$-th AP, $\textbf{G}_{mk}\in\mathbb{C}^{L_{mk,t}\times N}$, is given by
$\textbf{G}_{mk}=[\textbf{g}_{mk,1}(\textbf{t}_{m1}),...,\textbf{g}_{mk,N}(\textbf{t}_{mN})]$.
Similarly, the receive field-response vector for the channel between the $m$-th AP and the $k$-th user, $\textbf{f}_{mk}\left(\textbf{u}_k\right)\in\mathbb{C}^{L_{mk,r}\times 1}$ is expressed as
\begin{equation}
  \textbf{f}_{mk}\left(\textbf{u}_k\right)=[e^{j\frac{2\pi}{\lambda}\rho_{mk}^1(\textbf{u}_{k})},e^{j\frac{2\pi}{\lambda}\rho_{mk}^2(\textbf{u}_{k})},...,e^{j\frac{2\pi}{\lambda}\rho_{mk}^{L_{mk,r}}(\textbf{u}_{k})}]^T.\label{f_mk}
\end{equation}
Define the path-response matrix $\boldsymbol{\Sigma}_{mk}\in\mathbb{C}^{L_{mk,r}\times L_{mk,t}}$ as the response matrix between all the transmit and receive channels from $D_m$ to $O_k$. The entry in the $i$-th row and $j$-th column of $\boldsymbol{\Sigma}_{mk}$, i.e., $\boldsymbol{\Sigma}_{mk}(i,j)$, corresponds to the response coefficient between the $j$-th transmit path and the $i$-th receive path from the $m$-th AP to the $k$-th user\cite{11016053,11018493,10243545}. Then, the downlink channel vector
from the $m$-th AP and the $k$-th user, $\textbf{d}_{mk}\left(\textbf{u}_k\right)\in\mathbb{C}^{1\times N}$, is given by
\begin{equation}
\textbf{d}_{mk}\left(\textbf{u}_k\right)=\textbf{f}_{mk}^H\left(\textbf{u}_k\right)\boldsymbol{\Sigma}_{mk}\textbf{G}_{mk}.
\end{equation}

\subsection{Asynchronous Reception}
In practice, the asynchronous reception in the wireless transceiver can occur due to signal propagation delay differences caused by the varying positions of the distributed APs, leading to different propagation distances between each AP and a certain user\cite{10032129,8676341}. This results in unavoidable asynchronous signal arrivals and introduces delay phases, which can be multiplied in
the received signals \cite{10032129,10319677,9531352}. As a result, the delay phase can be introduced as
\begin{equation}
\begin{array}{ll}
   \displaystyle \vartheta_{mk}=e^{-j2\pi\frac{\Delta t_{mk}}{T_s}},
\end{array}
  \end{equation}
where $\Delta t_{mk}=\frac{d_{mk}-d_{m'k}}{c}=t_{mk}-t_{m'k}$ signifies the timing offset of the signal transmitted by the $m$-th AP and intended for the $k$-th user\cite{10804196,10319677,10319677}. Besides, $d_{mk}$ and $t_{mk}$ denote the propagation distance and time from
the $m$-th AP to the $k$-th user, respectively. $c$ is the speed of light and $T_s$ is the symbol duration. Without loss of generality, we define the $m'$-th AP as the reference AP for the $k$-th user, corresponding to the AP with the shortest propagation delay (i.e., the first arriving signal), with its relative timing offset $\Delta t_{m'k}=0$. 
In this work, we focus on a narrowband cell-free massive MIMO deployment in which the maximum timing offset between distributed APs and users is smaller than $T_s$. Under these conditions, the asynchronous reception effect primarily manifests as a phase shift. This deliberate scope allows us to avoid the confounding variable of inter-symbol interference (ISI) and to isolate exactly how FAS can exploit its spatial reconfigurability to reconfigure the physical channels constructively\cite{9531352,10032129,10008786}\footnote{Note that while the assumption of negligible ISI is highly applicable to narrowband scenarios, significant timing offsets in broadband scenarios (such as OFDM systems) can induce severe ISI when the asynchronous timing offset exceeds the CP length \cite{10319677,11100884}. Since incorporating ISI would require complex time-domain equalization that could obscure the foundational spatial DoFs unlocked by FASs, we defer the study of ISI to future investigation. Evaluating the impact of ISI and developing joint designs for FAS position optimization and advanced ISI mitigation techniques will be a critical direction for our future work to extend this framework to broadband systems \cite{10319677}.}. Thus, the efficient downlink channel from the $m$-th AP to the $k$-th user with delay phase is
  \begin{equation}
\begin{array}{ll}
   \displaystyle \textbf{h}_{mk}=\vartheta_{mk}\textbf{d}_{mk}\left(\textbf{u}_k\right).
\end{array}
  \end{equation}
The delay phase $\vartheta_{mk}$ is primarily determined by the positions of APs and users, making it essentially constant across multiple coherent blocks. Notably, this work assumes that the asynchronous reception arises mainly from the delay phase, and we will analyze hardware oscillator errors in our future work \cite{10008786,10032129}.

\section{Downlink Asynchronous Transmission and Performance Analysis}
\subsection{Coherent Downlink Asynchronous Transmission}
In this subsection, we assume that the cell-free massive MIMO system employs coherent transmission, in which each AP transmits the same data symbol to each user, as do the other APs \cite{10032129,8809413}. Accordingly,
$q_{k}\sim\mathcal{CN}(0,1)$ is the symbol intended for the $k$-th user, then the transmitted signal $\textbf{x}_m\in\mathbb{C}^{N\times 1}$ at the $m$-th AP can be given by
\begin{equation}
    \textbf{x}_m=\sum\nolimits_{k=1}^{K}\sqrt{\eta_{mk}p_d}{\textbf{v}}_{mk}q_{k},
\end{equation}
where the power control coefficient $\eta_{mk}$ satisfies $||\textbf{x}_m||^2 \leq p_d,~\forall m$, with the maximum downlink transmit power $p_d$ \cite{11196010}. Each AP adopts ${\textbf{v}}_{mk}\in\mathbb{C}^{N\times 1}$ as the transmit beamforming vector designated for the $k$-th user to precode the desired signal\cite{11196010}. 
Given that each user can only configure its FAS to a designated position, the received signal, $r_k$, at the $k$-th user, is given by
    \begin{equation}
\begin{array}{ll}
\vspace{2pt}
     \displaystyle r_k&\displaystyle=\sum\nolimits_{m=1}^M\textbf{h}_{mk}\textbf{x}_m+\textbf{n}_k\\&\displaystyle=
\underbrace{\sum\nolimits_{m=1}^M\sqrt{p_d\eta_{mk}}\textbf{h}_{mk}{\textbf{v}}_{mk}q_{k}}_{\text{desired signal}}\\&\displaystyle+
\underbrace{\sum\nolimits_{k'\neq k}\sum\nolimits_{m=1}^M\sqrt{p_d\eta_{mk'}}\textbf{h}_{mk}{\textbf{v}}_{mk'}q_{k'}}_{\text{interference}}+n_k,
\end{array}\label{r_k}
   \end{equation}
where $n_k\sim \mathcal{CN}(0,\sigma^2)$ is the additive white Gaussian noise. 
 We consider a practical scenario in which all FAS update their positions only in response to fading channel changes, rather than on a symbol-by-symbol basis. Thus, the selected antenna positions remain constant until the fading channels change, characterizing s-FAMA \cite{10146262,10066316}. We then define the SE of the $k$-th user for the coherent transmission as
 \begin{equation}
\begin{array}{ll}
   \displaystyle \text{SE}_k^\text{coh} = \text{log}_2\left(1+\frac{\displaystyle 
   p_d\big{|} \sum\nolimits_{m=1}^M\sqrt{\eta_{mk}}\textbf{h}_{mk}{\textbf{v}}_{mk} \big{|}^2
 }{\displaystyle
   p_d\sum\nolimits_{k'\neq k}\big{|}\sum\nolimits_{m=1}^M\sqrt{\eta_{mk'}}\textbf{h}_{mk}{\textbf{v}}_{mk'}\big{|}^2+\sigma^2
   }\right).
\end{array}\label{SE_k_renew}
  \end{equation}

In order to emphasize the deployment of FAS, we utilize low-complexity MR precoding relying on perfect channel estimates \cite{10146262,11018493} to derive analytical results \footnote{
This work assumes that perfect CSI is available at both the CPU and
APs\cite{11351331,11353414,11271835}. By assuming perfect CSI, we can theoretically quantify the spatial gains enabled by FAS, thereby isolating the asynchronous reception effect from the performance degradation caused by channel estimation errors. The performance results obtained herein serve as a baseline for measuring practical implementation losses.
We note that while existing efficient algorithms can accurately estimate CSI \cite {10755170,10497534}, practical wireless environments inevitably incur imperfect CSI, which degrades system performance. Investigating this performance degradation and designing robust optimization frameworks that account for estimation errors are vital next steps to demonstrate the robustness of FAS-enabled systems under realistic, dynamic conditions.
}. Then, the low-complexity delay phase forgotten (DF) MR precoding is characterized by $\textbf{v}_{mk}=\textbf{d}_{mk}^H\left(\textbf{u}_k\right)$ \cite{10032129,10008786}. The design of efficient downlink beamforming for asynchronous reception mitigation
is left for our future work.

\subsection{Non-Coherent Downlink Asynchronous Transmission}
This subsection employs non-coherent transmission, enabling each AP to transmit distinct data symbols to each user, to offset the phase synchronization requirements imposed by coherent transmission \cite{10032129,10804196,8809413}. Thus, 
$q_{mk}\sim\mathcal{CN}(0,1)$ is the symbol intended for the $k$-th user from the $m$-th AP, varying across different APs \cite{8809413}. Subsequently, the transmitted signal $\textbf{x}_m\in\mathbb{C}^{N\times 1}$ from the $m$-th AP 
is 
\begin{equation}
    \textbf{x}_m=\sum\nolimits_{k=1}^{K}\sqrt{\eta_{mk}p_d}{\textbf{v}}_{mk}q_{mk},
\end{equation}
then, the received signal, $r_k$ at the $k$-th user, is expressed as
 \begin{equation}
\begin{array}{ll}
\vspace{2pt}  \displaystyle r_k&\displaystyle=\sum\nolimits_{m=1}^M\textbf{h}_{mk}\textbf{x}_m+\textbf{n}_k\\&\displaystyle=\underbrace{\sum\nolimits_{m=1}^M\sqrt{p_d\eta_{mk}}\textbf{h}_{mk}{\textbf{v}}_{mk}q_{mk}}_{\text{desired signal}}\\&\displaystyle+
\underbrace{\sum\nolimits_{k'\neq k}\sum\nolimits_{m=1}^M\sqrt{p_d\eta_{mk'}}\textbf{h}_{mk}{\textbf{v}}_{mk'}q_{mk'}}_{\text{interference}}+n_k.
\end{array}\label{r_k_nc}
  \end{equation}
According to the received signal in \eqref{r_k_nc}, we propose that the $k$-th user utilizes the successive interference cancellation (with
arbitrary decoding order) outlined in \cite{8809413} to detect the $M$ signals. 
Similar to \eqref{SE_k_renew}, in the context of s-FAMA\cite{10146262,10066316}, the achievable downlink SE of the $k$-th user for non-coherent transmission can be re-written as 
 \begin{equation}
\begin{array}{ll}
   \displaystyle \text{SE}_k^\text{nc} = \text{log}_2\left(1+\frac{\displaystyle p_d\sum\nolimits_{m=1}^M\displaystyle 
   \big{|} \sqrt{\eta_{mk}}\textbf{h}_{mk}{\textbf{v}}_{mk} \big{|}^2
  }{\displaystyle p_d\sum\nolimits_{k'\neq k}\sum\nolimits_{m=1}^M\displaystyle
\big{|}\sqrt{\eta_{mk'}}\textbf{h}_{mk}{\textbf{v}}_{mk'}\big{|}^2+\sigma^2
   }\right).
\end{array}\label{SE_k_renew_nc}
  \end{equation}
  For non-coherent transmission, the emphasis is also placed on low-complexity DF-MR precoding $\textbf{v}_{mk}=\textbf{d}_{mk}^H\left(\textbf{u}_k\right)$. According to \eqref{SE_k_renew_nc}, it is evident that delay phases do not affect the SE performance of non-coherent transmission since there is no
cooperation among APs. Thus, non-coherent transmission can effectively counteract asynchronous reception effects, eliminating the delay phase but leading to inferior SE performance 
  \cite{10032129,10008786}. 

\subsection{Problem Formulation}
We aim to maximize the downlink sum SE of the proposed system in the presence of asynchronous reception by designing the
positions of the FAS and the downlink power control coefficients. Specifically, we can formulate the problem of maximizing
downlink sum SE as
\cite{10839251,9217298}
  \begin{subequations}
  \begin{align}
   P_1:~&\mathop {\max }\limits_{\{\textbf{u}_k\},~\{\eta_{mk}\}}~\text{SE}_{\text{sum}}=\sum\nolimits_{k=1}^K\text{SE}_k
   \\
   & \text{subject~to} \nonumber 
   \\
   &\textbf{u}_k\in \mathcal{C}_k,~\forall k,\label{position_constraint}\\
   & \sum\nolimits_{k=1}^K \eta_{mk}||\textbf{v}_{mk}||^2\leq1,~\forall m,\label{eta_constraint}
   \\
   &\eta_{mk}\geq0,~\forall m,k.\label{eta_constraint2}
  \end{align}
  \label{NAPGA_optimization}
 \end{subequations}
This joint optimization problem is recognized as non-convex\cite{11196010,9217298,11018493},
where \eqref{position_constraint} denotes the constraint on the feasible movement regions of FAS, \eqref{eta_constraint} and \eqref{eta_constraint2} represent the downlink power control coefficient constraints. Due to the coupling of the optimization variables
$\{\textbf{u}_k\}$ and $\{\eta_{mk}\}$
in the objective function, the optimization problem \eqref{NAPGA_optimization} is difficult to solve. Thus, an efficient algorithm based on the alternating
optimization (AO) framework is adopted, decomposing \eqref{NAPGA_optimization} into the following two sub-optimization problems:
   \begin{subequations}
  \begin{align}
   P_2:~&\mathop {\max }\limits_{\{\textbf{u}_k\}}~\text{SE}_{\text{sum}}=\sum\nolimits_{k=1}^K\text{SE}_k
   \\
   & \text{subject~to}~~ 
   \eqref{position_constraint}
  \end{align}
  \label{GA_optimization_MA}
 \end{subequations}
 and
  \begin{subequations}
  \begin{align}
   P_3:~&\mathop {\max }\limits_{\{\eta_{mk}\}}~\text{SE}_{\text{sum}}=\sum\nolimits_{k=1}^K\text{SE}_k
   \\
   & \text{subject~to}~~
    \eqref{eta_constraint}, \eqref{eta_constraint2}
  \end{align}
  \label{GA_optimization_power}
 \end{subequations}
 The sub-optimization problems \eqref{GA_optimization_MA} and \eqref{GA_optimization_power} are still known to be
non-convex and thus difficult to solve\cite{11196010,9217298,11018493}. Given the strong robustness and adaptability of the accelerated proximal gradient-based method \cite{NIPS2015_f7664060,9217298}, we propose the efficient NAPGA method based on the AO framework to tackle these sub-optimization problems and maximize the downlink sum SE \cite{NIPS2015_f7664060,10811786,10387264,xiu2025robustoptimizationmovableantennaaided}. The proposed NAPGA algorithm can achieve lower complexity, comparable performance, and faster convergence \cite{NIPS2015_f7664060,9217298}. The next section will introduce the efficient optimization algorithm in detail. 

\section{Downlink Sum SE Maximization based on Nonmonotone Accelerated Projected
Gradient Ascent Algorithm}
This section aims to solve the optimization problem $P_1$ in \eqref{NAPGA_optimization} by the proposed NAPGA method
to maximize the downlink sum SE. The processes of optimizing the FAS positions, the downlink power control coefficients, and the overall alternating optimization algorithm are given.

\subsection{Optimize $\textbf{u}_k$ given $\eta_{mk}$}
In the case of given $\eta_{mk},~\forall m,~k$, we first solve the FAS position sub-optimization problem $P_2$ in \eqref{GA_optimization_MA}. For solving $P_2$, we define $\textbf{u}$ as the vector of all FAS positions associated
with all users, given by $\textbf{u}=[\textbf{u}_1;\textbf{u}_2;...;\textbf{u}_K]\in\mathbb{C}^{2K\times1}$. Then, the downlink sum SE can be expressed as $g(\textbf{u})=\sum\nolimits_{k=1}^K\text{SE}_k(\textbf{u})$. To implement the proposed NAPGA algorithm, we can formulate $\nabla_{\textbf{u}}g(\textbf{u})$ as
\begin{equation}
\begin{array}{ll}
   \displaystyle 
   \nabla_{\textbf{u}}g(\textbf{u})=\left[\frac{\partial g(\textbf{u}) }{\partial\textbf{u}_1};\frac{\partial g(\textbf{u}) }{\partial\textbf{u}_2};...;\frac{\partial g(\textbf{u}) }{\partial\textbf{u}_K}\right]\in\mathbb{C}^{2K\times1}.
\end{array}\label{f_gradient}
  \end{equation}
Based on \eqref{f_mk}, we can first compute
$\frac{\displaystyle\partial\textbf{f}_{mk}(\textbf{u}_k)}{\displaystyle\partial\textbf{u}_k}$. Accordingly,
$\left[\frac{\displaystyle\partial\textbf{f}_{mk}(\textbf{u}_k)}{\displaystyle\partial\textbf{u}_k}\right]_1$, $\left[\frac{\displaystyle\partial\textbf{f}_{mk}(\textbf{u}_k)}{\displaystyle\partial\textbf{u}_k}\right]_2$ represent the elements of the first row and the second row of the gradient matrix of $\textbf{f}_{mk}(\textbf{u}_k)$, respectively, which are given by
 \begin{equation}
\begin{array}{ll}
\left[\frac{\displaystyle\partial\textbf{f}_{mk}(\textbf{u}_k)}{\displaystyle\partial\textbf{u}_k}\right]_1&\displaystyle=\frac{\displaystyle\partial\textbf{f}_{mk}(\textbf{u}_k)}{\displaystyle\partial x_k}\\&\displaystyle=j\frac{\displaystyle 2\pi}{\displaystyle \lambda}\text{diag}\left(\sin \theta_{mk,l}\cos \phi_{mk,l}\right)_1^{L_{mk,r}}\textbf{f}_{mk}(\textbf{u}_k),
   \end{array}
  \end{equation} 
   \begin{equation}
\begin{array}{ll}
\left[\frac{\displaystyle\partial\textbf{f}_{mk}(\textbf{u}_k)}{\displaystyle\partial\textbf{u}_k}\right]_2=\frac{\displaystyle\partial\textbf{f}_{mk}(\textbf{u}_k)}{\displaystyle\partial y_k}=j\frac{\displaystyle 2\pi}{\displaystyle \lambda}\text{diag}\left(\cos \theta_{mk,l}\right)_1^{L_{mk,r}}\textbf{f}_{mk}(\textbf{u}_k),
   \end{array}
  \end{equation} 
where $\text{diag}(a_l)_1^L$ represents a $L\times L$ diagonal matrix, in which the $l$-th element is $a_l$.

 \subsubsection{Coherent Transmission}For coherent transmission, 
we first define $\textbf{g}_{kk'}(\textbf{u})=\sum\nolimits_{m=1}^M\sqrt{\eta_{mk'}}\textbf{h}_{mk}{\textbf{v}}_{mk'},~\forall k, ~k'$; then, referring to \cite{11196010,9217298} and based on the chain rule, we can calculate $\frac{\displaystyle\partial g(\textbf{u}) }{\displaystyle\partial\textbf{u}_i}$ as
 \begin{equation}
\begin{array}{ll}
   \displaystyle 
   \frac{\partial g(\textbf{u}) }{\partial\textbf{u}_i}&\displaystyle=\frac{\partial}{\partial\textbf{u}_i}\sum\nolimits_{k=1}^K \text{log}_2\left(1+\frac{\displaystyle 
   p_d\big{|} \textbf{g}_{kk}(\textbf{u}) \big{|}^2
 }{\displaystyle
   p_d\sum\nolimits_{k'\neq k}\big{|}\textbf{g}_{kk'}(\textbf{u})\big{|}^2+\sigma^2
   }\right)\\&\displaystyle=\frac{1}{\text{ln}2}\left(\begin{array}{ll}\displaystyle
   \sum\nolimits_{k=1}^K
   
    \frac{
\frac{\displaystyle\partial}{\displaystyle\partial\textbf{u}_i}\left(\displaystyle
p_d\sum\nolimits_{k'=1}^K\big{|}\textbf{g}_{kk'}(\textbf{u})\big{|}^2+\sigma^2\right)
  }{\displaystyle
p_d\sum\nolimits_{k'=1}^K\big{|}\textbf{g}_{kk'}(\textbf{u})\big{|}^2+\sigma^2}
\\\displaystyle-\sum\nolimits_{k=1}^K
    \frac{
\frac{\displaystyle\partial}{\displaystyle\partial\textbf{u}_i}\left(\displaystyle
p_d\sum\nolimits_{k'\neq k}^K\big{|}\textbf{g}_{kk'}(\textbf{u})\big{|}^2+\sigma^2\right)
  }{\displaystyle
p_d\sum\nolimits_{k'\neq k}^K\big{|}\textbf{g}_{kk'}(\textbf{u})\big{|}^2+\sigma^2}\end{array}\right)
,
\end{array}\label{g_gradient_per_coh}
  \end{equation}
  where
 \begin{equation}
\begin{array}{ll}
   \displaystyle\frac{   \displaystyle\partial|\textbf{g}_{kk'}(\textbf{u})|^2}{   \displaystyle\partial\textbf{u}_i}&\displaystyle=\frac{   \displaystyle\partial\textbf{g}_{kk'}^H(\textbf{u})\textbf{g}_{kk'}(\textbf{u})}{   \displaystyle\partial\textbf{u}_i}
   \\&\displaystyle=
       2\mathfrak{RE}\left\{\textbf{g}_{kk'}^H(\textbf{u})\frac{   \displaystyle\partial\textbf{g}_{kk'}(\textbf{u})}{   \displaystyle\partial\textbf{u}_i}\right\},
\end{array}\label{g_kk_G}
  \end{equation}
and $\displaystyle\frac{   \displaystyle\partial\textbf{g}_{kk'}(\textbf{u})}{   \displaystyle\partial\textbf{u}_i}$ is given by \eqref{g_kk'_gradient} at the top of the next page.
   
 \begin{figure*}
  \begin{equation}
\begin{array}{ll}
   \displaystyle\displaystyle\frac{   \displaystyle\partial\textbf{g}_{kk'}(\textbf{u})}{   \displaystyle\partial\textbf{u}_i}&\displaystyle=\frac{   \displaystyle\partial\sum\nolimits_{m=1}^M\sqrt{\eta_{mk'}}\vartheta_{mk}\textbf{d}_{mk}(\textbf{u}_k)\textbf{d}_{mk'}^H(\textbf{u}_{k'})}{   \displaystyle\partial\textbf{u}_i}\displaystyle =\sum\nolimits_{m=1}^M\sqrt{\eta_{mk'}}\vartheta_{mk}\frac{   \displaystyle\partial\textbf{f}_{mk}^H\left(\textbf{u}_k\right)\boldsymbol{\Sigma}_{mk}\textbf{G}_{mk}\textbf{G}_{mk'}^H\boldsymbol{\Sigma}_{mk'}^H\textbf{f}_{mk'}\left(\textbf{u}_{k'}\right)}{   \displaystyle\partial\textbf{u}_i}
     \\&\displaystyle=\left\{\begin{array}{ll}
          \displaystyle
          2\sum\nolimits_{m=1}^M\sqrt{\eta_{mk}}\vartheta_{mk}\mathfrak{RE}\left\{\frac{   \displaystyle\partial\textbf{f}_{mk}^H\left(\textbf{u}_k\right)}{   \displaystyle\partial\textbf{u}_k}\boldsymbol{\Sigma}_{mk}\textbf{G}_{mk}\textbf{d}_{mk}\left(\textbf{u}_k\right)
   \right\},~i=k=k'\\
   \displaystyle \sum\nolimits_{m=1}^M\sqrt{\eta_{mk'}}\vartheta_{mk}\frac{   \displaystyle\partial\textbf{f}_{mk}^H\left(\textbf{u}_k\right)}{   \displaystyle\partial\textbf{u}_k}\boldsymbol{\Sigma}_{mk}\textbf{G}_{mk}\textbf{d}_{mk'}\left(\textbf{u}_{k'}\right)
,~i=k,~i\neq k'\\

\displaystyle \sum\nolimits_{m=1}^M\sqrt{\eta_{mk'}}\vartheta_{mk}\frac{   \displaystyle\partial\textbf{f}_{mk'}^H\left(\textbf{u}_{k'}\right)}{   \displaystyle\partial\textbf{u}_{k'}}\boldsymbol{\Sigma}_{mk'}\textbf{G}_{mk'}\textbf{d}_{mk}\left(\textbf{u}_{k}\right)
,~i=k',~i\neq k
\\
0,~\text{otherwise}
     \end{array}
     
     \right..
\end{array}\label{g_kk'_gradient}
  \end{equation}
\hrulefill
\vspace{-5pt}
   \end{figure*}

   \subsubsection{Non-Coherent Transmission}For non-coherent transmission, 
 we can define $\textbf{b}_{mkk'}(\textbf{u})=\textbf{h}_{mk}{\textbf{v}}_{mk'}~\forall m,~ k, ~k'$; then, referring to \eqref{SE_k_renew_nc}, we can calculate $\frac{\displaystyle\partial g(\textbf{u}) }{\displaystyle\partial\textbf{u}_i}$ as
  \begin{align}
\displaystyle 
   &\displaystyle\frac{\partial g(\textbf{u}) }{\partial\textbf{u}_i}\nonumber\\&\displaystyle=\frac{\partial}{\partial\textbf{u}_i}\sum\limits_{k=1}^K \text{log}_2\left(1+\frac{\displaystyle 
   p_d\sum\nolimits_{m=1}^M\big{|} \sqrt{\eta_{mk}}\textbf{b}_{mkk}(\textbf{u}) \big{|}^2
 }{\displaystyle
   p_d\sum\nolimits_{k'\neq k}\sum\nolimits_{m=1}^M\big{|} \sqrt{\eta_{mk'}}\textbf{b}_{mkk'}(\textbf{u}) \big{|}^2+\sigma^2
   }\right)\nonumber\\&\displaystyle=\frac{1}{\text{ln}2}\left(\begin{array}{ll}\displaystyle\sum\limits_{k=1}^K
    \frac{
\frac{\displaystyle\partial}{\displaystyle\partial\textbf{u}_i}\left(\displaystyle
p_d\sum\nolimits_{k'=1}^K\sum\nolimits_{m=1}^M\big{|} \sqrt{\eta_{mk'}}\textbf{b}_{mkk'}(\textbf{u}) \big{|}^2+\sigma^2\right)
  }{\displaystyle
p_d\sum\nolimits_{k'=1}^K\sum\nolimits_{m=1}^M\big{|} \sqrt{\eta_{mk'}}\textbf{b}_{mkk'}(\textbf{u}) \big{|}^2+\sigma^2}
\\\displaystyle-\sum\limits_{k=1}^K
    \frac{
\frac{\displaystyle\partial}{\displaystyle\partial\textbf{u}_i}\left(\displaystyle
p_d\sum\nolimits_{k'\neq k}^K\sum\nolimits_{m=1}^M\big{|} \sqrt{\eta_{mk'}}\textbf{b}_{mkk'}(\textbf{u}) \big{|}^2+\sigma^2\right)
  }{\displaystyle
p_d\sum\nolimits_{k'\neq k}^K\sum\nolimits_{m=1}^M\big{|} \sqrt{\eta_{mk'}}\textbf{b}_{mkk'}(\textbf{u}) \big{|}^2+\sigma^2}\end{array}\right)
,\label{g_gradient_per}   
  \end{align}
  where
\begin{equation}
\begin{array}{ll}
   \displaystyle\frac{   \displaystyle\partial|\textbf{b}_{mkk'}(\textbf{u})|^2}{   \displaystyle\partial\textbf{u}_i}&\displaystyle=\frac{   \displaystyle\partial\textbf{b}_{mkk'}^H(\textbf{u})\textbf{b}_{mkk'}(\textbf{u})}{   \displaystyle\partial\textbf{u}_i}
   \\&\displaystyle=
        2\mathfrak{RE}\left\{\textbf{b}_{mkk'}^H(\textbf{u})\frac{   \displaystyle\partial\textbf{b}_{mkk'}(\textbf{u})}{   \displaystyle\partial\textbf{u}_i}\right\},
\end{array}\label{b_mkk_G}
  \end{equation}
     \begin{equation}
\begin{array}{ll}
   \displaystyle&\displaystyle\frac{   \displaystyle\partial\textbf{b}_{mkk'}(\textbf{u})}{   \displaystyle\partial\textbf{u}_i}\displaystyle=\frac{   \displaystyle\partial\vartheta_{mk}\textbf{d}_{mk}(\textbf{u}_k)\textbf{d}_{mk'}^H(\textbf{u}_{k'})}{   \displaystyle\partial\textbf{u}_i}
     \\&\displaystyle =\left\{
     \begin{array}{ll}
          2\vartheta_{mk}\mathfrak{RE}\left\{\frac{   \displaystyle\partial\textbf{f}_{mk}^H\left(\textbf{u}_k\right)}{   \displaystyle\partial\textbf{u}_k}\boldsymbol{\Sigma}_{mk}\textbf{G}_{mk}\textbf{d}_{mk}\left(\textbf{u}_k\right)
   
   \right\},~i=k=k' \\
         \vartheta_{mk}\frac{   \displaystyle\partial\textbf{f}_{mk}^H\left(\textbf{u}_k\right)}{   \displaystyle\partial\textbf{u}_k}\boldsymbol{\Sigma}_{mk}\textbf{G}_{mk}\textbf{d}_{mk'}\left(\textbf{u}_{k'}\right),~i=k,~i\neq k'\\
         \vartheta_{mk}\frac{   \displaystyle\partial\textbf{f}_{mk'}^H\left(\textbf{u}_{k'}\right)}{   \displaystyle\partial\textbf{u}_{k'}}\boldsymbol{\Sigma}_{mk'}\textbf{G}_{mk'}\textbf{d}_{mk}\left(\textbf{u}_{k}\right),~i=k',~i\neq k\\
         0,~\text{otherwise}
     \end{array}
     \right..
\end{array}\label{b_mkk'_gradient}
  \end{equation}
  \subsubsection{Projection on $\mathcal{C}$}
  To guarantee the constraint in \eqref{position_constraint}, each update for FAS positions is followed by a projection operation $\text{P}_{\mathcal{C}}\left(\textbf{u}\right)$, which is defined as
\begin{equation}
  \begin{array}{cc}
[\text{P}_{\mathcal{C}}\left(\textbf{u}_k\right)]_{p}=\displaystyle\left\{\begin{array}{ll}
d_\text{min},~\text{if} ~[\textbf{u}_k]_p<d_\text{min}
\\
{[}\textbf{u}_k{]}_p,~\text{if}~ d_\text{min}<[\textbf{u}_k]_p<d_\text{max}
\\
d_\text{max},~\text{if}~ [\textbf{u}_k]_p>d_\text{max}
\end{array}
\right.,~\forall k,
\end{array}
\end{equation}
where $[\cdot]_p$ denotes the $p$-th entry of the argument. Note that the convergence behavior of the gradient ascent method highly depends on the step size, which can be calculated by the backtracking line search\cite{hu2025uplinktransmissiondesignfluid,10388242,11098906}. Specifically, the step size iteratively decreases during the search process with a scaling factor $\xi\in\left(0,1\right)$ and a control parameter $\varpi\in\left(0,1\right)$, until it satisfies the Armijo-Goldstein condition\cite{10811786,hu2025uplinktransmissiondesignfluid}. Then, the step size design for FAS position optimization follows Algorithm \ref{Step_size_search}. 
 \begin{algorithm}[t!]
   \caption{Backtracking Line Search for Step Size} 
\begin{algorithmic}[1]
\renewcommand{\algorithmicrequire}{\textbf{Inputs:}}
\Require
$\xi\in(0,1)$, $\varpi\in(0,1)$, $\alpha>0$
\renewcommand{\algorithmicensure}{\textbf{Output:}}
\Ensure
 $\alpha_1^{(n+1)}$, $\alpha_2^{(n+1)}$
\State Initialize the step size as $\alpha_1^{(n+1)}=\alpha$, $\alpha_2^{(n+1)}=\alpha$
\Repeat 
\State $\textbf{v}^{n+1}=P_{\mathcal{C}}\left(\textbf{u}^n+\alpha_1^{(n+1)}\nabla_{\textbf{u}^n} g(\textbf{u}^n)\right)$ for Algorithm \ref{Algorithm_NAPG_MA}
\State OR
\State $\textbf{v}^{n+1}=P_{\mathcal{S}}\left(\boldsymbol{\mu}^n+\alpha_1^{(n+1)}\nabla_{\boldsymbol{\mu}^n} f(\boldsymbol{\mu}^n)\right)$ for Algorithm \ref{Algorithm_NAPG_power_control}
\State $\alpha_1^{(n+1)}=\xi \alpha_1^{(n+1)}$
\Until {$g(\textbf{v}^{n+1})\geq c_n+\varpi\alpha_1^{(n+1)}||\textbf{v}^{n+1}-\textbf{u}^n||$} for Algorithm \ref{Algorithm_NAPG_MA} OR {$f(\textbf{v}^{n+1})\geq c_n+\varpi\alpha_1^{(n+1)}||\textbf{v}^{n+1}-\boldsymbol{\mu}^n||$} for Algorithm \ref{Algorithm_NAPG_power_control} 
\Repeat 
\State $\textbf{z}^{n+1}=P_{\mathcal{C}}\left(\textbf{y}^n+\alpha_2^{(n+1)}\nabla_{\textbf{u}^n}  g(\textbf{y}^n)\right)$ for Algorithm \ref{Algorithm_NAPG_MA}
\State OR
\State $\textbf{z}^{n+1}=P_{\mathcal{S}}\left(\textbf{y}^n+\alpha_2^{(n+1)}\nabla_{\boldsymbol{\mu}^n}  f(\textbf{y}^n)\right)$ for Algorithm \ref{Algorithm_NAPG_power_control}
\State $\alpha_2^{(n+1)}=\xi \alpha_2^{(n+1)}$
\Until {$g(\textbf{z}^{n+1})\geq g(\textbf{y}^{n})+\varpi\alpha_2^{(n+1)}||\textbf{z}^{n+1}-\textbf{y}^{n}||$} for Algorithm \ref{Algorithm_NAPG_MA} OR {$f(\textbf{z}^{n+1})\geq f(\textbf{y}^{n})+\varpi\alpha_2^{(n+1)}||\textbf{z}^{n+1}-\textbf{y}^{n}||$} for Algorithm \ref{Algorithm_NAPG_power_control} 
\end{algorithmic}
\label{Step_size_search}
 \end{algorithm} 
 \begin{algorithm}[t!]
   \caption{NAPGA-Based FAS Position Optimization} 
\begin{algorithmic}[1]
\renewcommand{\algorithmicrequire}{\textbf{Inputs:}}
\Require
$\epsilon$ (tolerance), IterMax; $\textbf{u}^0\geq 0$, $t_0=t_1=1$, $q_1=1$, $\delta\in(0,1)$, $\nu\in[0,1)$
\renewcommand{\algorithmicensure}{\textbf{Output:}}
\Ensure
$\textbf{u}$
\State Initialize $\textbf{u}^1=\textbf{z}^1=\textbf{u}^0$, $c_1=g(\textbf{u}^1)$
\For{$\text{n}= 1 :\text{IterMax}$}
\State $\textbf{y}^n=\textbf{u}^n+\frac{\displaystyle t_{n-1}}{\displaystyle t_n}(\textbf{z}^n-\textbf{u}^n)+\frac{\displaystyle t_{n-1}-1}{\displaystyle t_n}(\textbf{u}^n-\textbf{u}^{n-1})$
\State Obtain the step size based on Algorithm \ref{Step_size_search}
\State $\textbf{v}^{n+1}=P_{\mathcal{C}}(\textbf{u}^n+\alpha_1^{(n+1)}\nabla g(\textbf{u}^n))$
\State $\textbf{z}^{n+1}=P_{\mathcal{C}}(\textbf{y}^n+\alpha_2^{(n+1)}\nabla g(\textbf{y}^n))$
\If{$g(\textbf{z}^{n+1})\geq c_n+\delta||\textbf{z}^{n+1}-\textbf{y}^{n}||^2$}
\State $\textbf{u}^{n+1}=\textbf{z}^{n+1}$
 \Else 
 \State $\textbf{u}^{n+1}=\left\{\begin{array}{ll}
  \textbf{z}^{n+1},~g(\textbf{z}^{n+1})>g(\textbf{v}^{n+1})
  \\
  \textbf{v}^{n+1},~\text{otherwise}
 \end{array}\right.$
 \EndIf
\State $t_{n+1}=\frac{\displaystyle 1+\sqrt{4t_n^2+1}}{\displaystyle 2}$
\State $q_{n+1}=\nu q_n+1$
\State $c_{n+1}=\frac{\displaystyle\nu q_nc_n+g(\text{u}^{n+1})}{\displaystyle q_{n+1}}$
\If {$\big{|}g(\textbf{u}^{n+1})-g(\textbf{u}^{n})\big{|}\leq \epsilon$}
\State $\textbf{break}$
\EndIf
\EndFor
\end{algorithmic}
\label{Algorithm_NAPG_MA}
 \end{algorithm}
\subsubsection{Optimization Algorithm and Complexity Analysis}
 Specifically, we illustrate
the iterative procedure of the NAPGA algorithm in Algorithm \ref{Algorithm_NAPG_MA}, aiming to converge to a stationary point of the sub-optimization problem \eqref{GA_optimization_MA}. 
The computational complexity is expressed using big-$\mathcal{O}$ notation. It is clear that the complexity of finding $\nabla g(\textbf{u})$ is $\mathcal{O}(MK^2N+MKNL^2)$, where $L$ is the number of receive paths with $L_{mk,r}=L,~\forall m,k$. The complexity of projection is $\mathcal{O}(K)$. Then, the complexity to solve \eqref{GA_optimization_MA} is $\mathcal{O}(T_1(MK^2N+MKNL^2+T_{inner,\textbf{u}}MK^2N))$, where $T_1$ is the number of iterations to complete Algorithm \ref{Algorithm_NAPG_MA} and $\mathcal{O}(T_{inner,\textbf{u}}MK^2N)$ is the complexity to find a feasible step size for FAS position optimization, where $T_{inner,\textbf{u}}$ is the average number of backtracking steps per iteration.

  \subsection{Optimize $\eta_{mk}$ given $\textbf{u}_k$}\label{PW_opt}

In the case of given $\textbf{u}_k,~\forall k$, we solve the power control sub-optimization problem $P_3$ in \eqref{GA_optimization_power}. From Section III, we note that $||\textbf{x}_m||^2 =p_d\sum\nolimits_{k=1}^K\eta_{mk}||\textbf{v}_{mk}||^2\leq p_d,~\forall m$; then, we can define $w_{mk}=||\textbf{v}_{mk}||^2$ and obtain $\sum\nolimits_{k=1}^K\eta_{mk}w_{mk}\leq 1,~\forall m$. To facilitate the proposed optimization algorithm, we define ${\boldsymbol{\mu}}_m$ as the vector of all power control coefficients associated
with the $m$-th AP with ${\boldsymbol{\mu}}_m=[\sqrt{\eta_{m1}w_{m1}};\sqrt{\eta_{m2}w_{m2}};...;\sqrt{\eta_{mK}w_{mK}}]\in\mathbb{C}^{K\times1}$ and $||\boldsymbol{\mu}_m||^2\leq1$. Therefore, the feasible region is $\mathcal{S}=\left\{\boldsymbol{\mu}|\boldsymbol{\mu}\geq0;||\boldsymbol{\mu}_m||^2\leq1,~\forall m\right\}$. If we define $\boldsymbol{\mu}=[\boldsymbol{\mu}_1;\boldsymbol{\mu}_2;...;\boldsymbol{\mu}_M]\in\mathbb{C}^{MK\times1}$, the downlink sum SE can be expressed as $f(\boldsymbol{\mu})=\sum\nolimits_{k=1}^K\text{SE}_k(\boldsymbol{\mu})$. To implement the proposed NAPGA algorithm, we first need to compute $\displaystyle 
\nabla_{\boldsymbol{\mu}} f(\boldsymbol{\mu})$, which is given by
\begin{equation}
\begin{array}{ll}
   \displaystyle 
   \nabla_{\boldsymbol{\mu}} f(\boldsymbol{\mu})=\left[\frac{\partial f(\boldsymbol{\mu}) }{\partial\boldsymbol{\mu}_1};\frac{\partial f(\boldsymbol{\mu}) }{\partial\boldsymbol{\mu}_2};...;\frac{\partial f(\boldsymbol{\mu}) }{\partial\boldsymbol{\mu}_K}\right]\in\mathbb{C}^{MK\times1}.
\end{array}\label{f_gradient}
  \end{equation}

  \begin{algorithm}[t!]
   \caption{NAPGA-Based Downlink Power Control} 
\begin{algorithmic}[1]
\renewcommand{\algorithmicrequire}{\textbf{Inputs:}}
\Require
$\epsilon$ (tolerance), IterMax; $\boldsymbol{\mu}^0\geq 0$, $t_0=t_1=1$, $q_1=1$, $\delta\in(0,1)$, $\nu\in[0,1)$
\renewcommand{\algorithmicensure}{\textbf{Output:}}
\Ensure
$\boldsymbol{\mu}$
\State Initialize $\boldsymbol{\mu}^1=\textbf{z}^1=\boldsymbol{\mu}^0$,  $c_1=f(\boldsymbol{\mu}^1)$
\For{$\text{n}= 1 :\text{IterMax}$}
\State $\textbf{y}^n=\boldsymbol{\mu}^n+\frac{\displaystyle t_{n-1}}{\displaystyle t_n}(\textbf{z}^n-\boldsymbol{\mu}^n)+\frac{\displaystyle t_{n-1}-1}{\displaystyle t_n}(\boldsymbol{\mu}^n-\boldsymbol{\mu}^{n-1})$
\State Obtain the step size based on Algorithm \ref{Step_size_search}
\State $\textbf{v}^{n+1}=P_{\mathcal{S}}\left(\boldsymbol{\mu}^n+\alpha_1^{(n+1)}\nabla f(\boldsymbol{\mu}^n)\right)$
\State $\textbf{z}^{n+1}=P_{\mathcal{S}}\left(\textbf{y}^n+\alpha_2^{(n+1)}\nabla  f(\textbf{y}^n)\right)$
\If {$f(\textbf{z}^{n+1})\geq c_n+\delta||\textbf{z}^{n+1}-\textbf{y}^{n}||^2$}
\State $\boldsymbol{\mu}^{n+1}=\textbf{z}^{n+1}$
 \Else 
\State $\boldsymbol{\mu}^{n+1}=\left\{\begin{array}{ll}
  \textbf{z}^{n+1},~f(\textbf{z}^{n+1})>f(\textbf{v}^{n+1})
  \\
  \textbf{v}^{n+1},~\text{otherwise}
 \end{array}\right.$
 \EndIf
\State $t_{n+1}=\frac{\displaystyle 1+\sqrt{4t_n^2+1}}{\displaystyle 2}$
\State $q_{n+1}=\nu q_n+1$
\State $c_{n+1}=\frac{\displaystyle\nu q_nc_n+f(\boldsymbol{\mu}^{n+1})}{\displaystyle q_{n+1}}$
\If {$\big{|}f(\boldsymbol{\mu}^{n+1})-f(\boldsymbol{\mu}^{n})\big{|}\leq \epsilon$}
\State $\textbf{break}$
\EndIf
\EndFor
\end{algorithmic}
\label{Algorithm_NAPG_power_control}
 \end{algorithm} 
\subsubsection{Coherent Transmission}For coherent transmission, the downlink SE of the $k$-th user in \eqref{SE_k_renew} can be rewritten as
 \begin{equation}
\begin{array}{ll}
   \displaystyle \text{SE}_k = \text{log}_2\left(1+\frac{\displaystyle 
   p_d\big{|} \textbf{D}_{kk}^T\boldsymbol{\mu}_k \big{|}^2
 }{\displaystyle
   p_d\sum\nolimits_{k'\neq k}\big{|}\textbf{D}_{kk'}^T{\boldsymbol{\mu}}_{k'}\big{|}^2+\sigma^2
   }\right),
\end{array}\label{SE_k_coh}
  \end{equation}
  where $\boldsymbol{\mu}_k=[\mu_{1k};\mu_{2k};...;\mu_{Mk}]\in\mathbb{C}^{M\times1}$ and $\displaystyle
\textbf{D}_{kk'}=\left[\frac{\textbf{h}_{1k}{\textbf{v}}_{1k'}}{\sqrt{w_{1k'}}};\frac{\textbf{h}_{2k}{\textbf{v}}_{2k'}}{\sqrt{w_{2k'}}};...;\frac{\textbf{h}_{Mk}{\textbf{v}}_{Mk'}}{\sqrt{w_{Mk'}}}\right]\in\mathbb{C}^{M\times1}$. 
Referring to \eqref{g_gradient_per_coh}, we can calculate $\displaystyle 
   \frac{\partial f(\boldsymbol{\mu}) }{\partial\boldsymbol{\mu}_i}$  as \eqref{f_gradient_per} at the top of this page 
   \begin{figure*}
\begin{equation}
\begin{array}{ll}
   \displaystyle 
   \frac{\partial f(\boldsymbol{\mu}) }{\partial\boldsymbol{\mu}_i}&\displaystyle=\frac{\partial}{\partial\boldsymbol{\mu}_i}\sum\nolimits_{k=1}^K \text{log}_2\left(1+\frac{\displaystyle 
   p_d\big{|} \textbf{D}_{kk}^T\boldsymbol{\mu}_k \big{|}^2
 }{\displaystyle
   p_d\sum\nolimits_{k'\neq k}\big{|}\textbf{D}_{kk'}^T{\boldsymbol{\mu}}_{k'}\big{|}^2+\sigma^2
   }\right)\\&\displaystyle=\displaystyle\frac{1}{\text{ln}2}\left(
\displaystyle\sum\nolimits_{k=1}^K
    \frac{
\frac{\displaystyle\partial}{\displaystyle\partial\boldsymbol{\mu}_i}\left(\displaystyle
p_d\sum\nolimits_{k'=1}^K\big{|}\textbf{D}_{kk'}^T{\boldsymbol{\mu}}_{k'}\big{|}^2+\sigma^2\right)
  }{\displaystyle
p_d\sum\nolimits_{k'=1}^K\big{|}\textbf{D}_{kk'}^T{\boldsymbol{\mu}}_{k'}\big{|}^2+\sigma^2}
-\sum\nolimits_{k=1}^K
    \frac{
\frac{\displaystyle\partial}{\displaystyle\partial\boldsymbol{\mu}_i}\left(\displaystyle
p_d\sum\nolimits_{k'\neq k}^K\big{|}\textbf{D}_{kk'}^T{\boldsymbol{\mu}}_{k'}\big{|}^2+\sigma^2\right)
  }{\displaystyle
p_d\sum\nolimits_{k'\neq k}^K\big{|}\textbf{D}_{kk'}^T{\boldsymbol{\mu}}_{k'}\big{|}^2+\sigma^2}   \right)

\\&\displaystyle=\displaystyle\frac{1}{\text{ln}2}\left(
\displaystyle\sum\nolimits_{k=1}^K
    \frac{\displaystyle
2p_d\mathfrak{RE}\{\textbf{D}_{ki}^*\textbf{D}_{ki}^T{\boldsymbol{\mu}}_{i}\}}{\displaystyle
p_d\sum\nolimits_{k'=1}^K\big{|}\textbf{D}_{kk'}^T{\boldsymbol{\mu}}_{k'}\big{|}^2+\sigma^2}
- \underbrace{\sum\nolimits_{k=1}^K
  \frac{\displaystyle
2p_d\mathfrak{RE}\{\textbf{D}_{ki}^*\textbf{D}_{ki}^T{\boldsymbol{\mu}}_{i}\}}{\displaystyle
p_d\sum\nolimits_{k'\neq k}^K\big{|}\textbf{D}_{kk'}^T{\boldsymbol{\mu}}_{k'}\big{|}^2+\sigma^2}}_{i\neq k}  \right).
\end{array}\label{f_gradient_per}
  \end{equation}
  \hrulefill
   \vspace{-5pt}
  \end{figure*}
with
 \begin{equation}
\begin{array}{ll}
\frac{\displaystyle\partial}{\displaystyle\partial\boldsymbol{\mu}_i}p_d\big{|} \textbf{D}_{kk'}^T\boldsymbol{\mu}_{k'} \big{|}^2&\displaystyle=p_d\frac{   \displaystyle\partial
\boldsymbol{\mu}_{k'} ^T\textbf{D}_{kk'}^*\textbf{D}_{kk'}^T\boldsymbol{\mu}_{k'}  
}{   \displaystyle\partial{\boldsymbol{\mu}_i}}\\&\displaystyle=\left\{\begin{array}{ll}
2p_d\mathfrak{RE}\{\textbf{D}_{ki}^*\textbf{D}_{ki}^T\boldsymbol{\mu}_i\},~k'=i\\
0,~i\neq k
\end{array}\right.,
   \end{array}
  \end{equation}
  \subsubsection{Non-Coherent Transmission}For non-coherent transmission, the SE of the $k$-th user in \eqref{SE_k_renew_nc} can be rewritten as
 \begin{equation}
\begin{array}{ll}
   \displaystyle \text{SE}_k = \text{log}_2\left(1+\frac{\displaystyle 
   p_d\boldsymbol{\mu}_k^T\bar{\textbf{D}}_{kk}\boldsymbol{\mu}_k 
 }{\displaystyle
   p_d\sum\nolimits_{k'\neq k}\boldsymbol{\mu}_{k'}^T\bar{\textbf{D}}_{kk'}\boldsymbol{\mu}_{k'} +\sigma^2
   }\right),
\end{array}\label{SE_k_nonco}
  \end{equation}
   where $\displaystyle
\bar{\textbf{D}}_{kk'}=\text{diag}\left(\left[\frac{|\textbf{h}_{1k}{\textbf{v}}_{1k'}|^2}{w_{1k'}};\frac{|\textbf{h}_{2k}{\textbf{v}}_{2k'}|^2}{w_{2k'}};...;\frac{|\textbf{h}_{Mk}{\textbf{v}}_{Mk'}|^2}{w_{Mk'}}\right]\right)\in\mathbb{C}^{M\times M}$. Similar to \eqref{f_gradient_per}, we can calculate $\displaystyle 
   \frac{\partial f(\boldsymbol{\mu}) }{\partial\boldsymbol{\mu}_i}$ for the non-coherent transmission as \eqref{f_gradient_per_nc} at the top of the next page
   \begin{figure*}
\begin{equation}
\begin{array}{ll}
   \displaystyle 
   \frac{\partial f(\boldsymbol{\mu}) }{\partial\boldsymbol{\mu}_i}&\displaystyle=\frac{\partial}{\partial\boldsymbol{\mu}_i}\sum\nolimits_{k=1}^K \text{log}_2\left(1+\frac{\displaystyle 
   p_d\boldsymbol{\mu}_k^T\bar{\textbf{D}}_{kk}\boldsymbol{\mu}_k 
 }{\displaystyle
   p_d\sum\nolimits_{k'\neq k}\boldsymbol{\mu}_{k'}^T\bar{\textbf{D}}_{kk'}\boldsymbol{\mu}_{k'} +\sigma^2
   }\right)\\&\displaystyle=\displaystyle\frac{1}{\text{ln}2}\left(
\displaystyle\sum\nolimits_{k=1}^K
   
    \frac{
\frac{\displaystyle\partial}{\displaystyle\partial\boldsymbol{\mu}_i}\left(\displaystyle
p_d\sum\nolimits_{k'=1}^K\boldsymbol{\mu}_{k'}^T\bar{\textbf{D}}_{kk'}\boldsymbol{\mu}_{k'} +\sigma^2\right)
  }{\displaystyle
p_d\sum\nolimits_{k'=1}^K\boldsymbol{\mu}_{k'}^T\bar{\textbf{D}}_{kk'}\boldsymbol{\mu}_{k'} +\sigma^2}
-\sum\nolimits_{k=1}^K
    \frac{
\frac{\displaystyle\partial}{\displaystyle\partial\boldsymbol{\mu}_i}\left(\displaystyle
p_d\sum\nolimits_{k'\neq k}^K\boldsymbol{\mu}_{k'}^T\bar{\textbf{D}}_{kk'}\boldsymbol{\mu}_{k'} +\sigma^2\right)
  }{\displaystyle
p_d\sum\nolimits_{k'\neq k}^K\boldsymbol{\mu}_{k'}^T\bar{\textbf{D}}_{kk'}\boldsymbol{\mu}_{k'} +\sigma^2}\right)
\\&\displaystyle=\displaystyle\frac{1}{\text{ln}2}\left(
\displaystyle\sum\nolimits_{k=1}^K
   
    \frac{
2p_d\bar{\textbf{D}}_{ki}\boldsymbol{\mu}_i }{\displaystyle
p_d\sum\nolimits_{k'=1}^K\boldsymbol{\mu}_{k'}^T\bar{\textbf{D}}_{kk'}\boldsymbol{\mu}_{k'} +\sigma^2}
- \underbrace{\sum\nolimits_{k=1}^K
   \frac{
2p_d\bar{\textbf{D}}_{ki}\boldsymbol{\mu}_i}{\displaystyle
p_d\sum\nolimits_{k'\neq k}^K\boldsymbol{\mu}_{k'}^T\bar{\textbf{D}}_{kk'}\boldsymbol{\mu}_{k'} +\sigma^2}}_{i\neq k}\right)
.
\end{array}\label{f_gradient_per_nc}
  \end{equation}
  \hrulefill
   \vspace{-0pt}
     \end{figure*}
  with
\begin{equation}
\begin{array}{ll}
\frac{\displaystyle\partial}{\displaystyle\partial\boldsymbol{\mu}_i}p_d\boldsymbol{\mu}_{k'}^T\bar{\textbf{D}}_{kk'}\boldsymbol{\mu}_{k'}=\left\{\begin{array}{ll}
2p_d\bar{\textbf{D}}_{ki}\boldsymbol{\mu}_i,~k'=i\\
0,~i\neq k
\end{array}\right..
   \end{array}
  \end{equation}
\subsubsection{Projection on $\mathcal{S}$}
To meet the constraint in \eqref{eta_constraint}, \eqref{eta_constraint2}, we introduce the projection $P_{\mathcal{S}}(\textbf{x})$, which is the solution of 
\begin{equation}
\begin{array}{ll}
\displaystyle \mathop {\text{minimize}}\limits_{\boldsymbol{\mu}\in\mathbb{R}^{MK\times 1}}\left\{||\boldsymbol{\mu}-\textbf{x}||^2~|~\boldsymbol{\mu}\geq0,~||\boldsymbol{\mu}_m||^2<1,~\forall m\right\},
   \end{array}
   \label{P_S}
  \end{equation}
which can be decomposed as
\begin{equation}
\begin{array}{ll}
\displaystyle \mathop {\text{minimize}}\limits_{\boldsymbol{\mu}_m\in\mathbb{R}^{K\times 1}}\left\{||\boldsymbol{\mu}_m-\textbf{x}_m||^2~|~\boldsymbol{\mu}_m\geq0,~||\boldsymbol{\mu}_m||^2<1,~\forall m\right\}.
   \end{array}
   \label{P_S2}
  \end{equation}
The above problem can be treated as the projection onto the intersection of a ball and the positive orthant. In this case, the analytical solution can be given by\cite{9217298,9709200}
  \begin{equation}
\begin{array}{ll}
\displaystyle\boldsymbol{\mu}_m=\displaystyle\frac{\displaystyle 1}{\displaystyle\text{max}\left(1,||[\textbf{x}_m]_{+}||\right)}[\textbf{x}_m]_{+},
   \end{array}
   \label{P_S_solution}
  \end{equation}
where $[\textbf{x}]_{+}$ denotes the projector onto
the positive orthant\cite{9217298,9709200}. Similar to the FAS position design process, the tailored backtracking line search for the proposed NAPGA algorithm is given in Algorithm \ref{Step_size_search}.

\subsubsection{Optimization Algorithm and Complexity Analysis}
According to the above-mentioned observations, we illustrate
the iterative procedure of the NAPGA algorithm in Algorithm \ref{Algorithm_NAPG_power_control}, aiming to converge to a stationary point of the sub-optimization problem \eqref{GA_optimization_power}. The calculation of the static inner products $\textbf{D}_{kk'},~\forall k,~k'$ takes 
$\mathcal{O}(MNK^2)$. Then, the complexity of finding $\nabla f(\boldsymbol{\mu})$ is $\mathcal{O}(MK^2)$. The complexity of projection is $\mathcal{O}(MK)$. Therefore, the complexity to solve \eqref{GA_optimization_power} is $\mathcal{O}(MNK^2+T_2(1+T_{inner,\boldsymbol{\mu}})MK^2)$, where $T_2$ is the number of iterations needed to complete Algorithm \ref{Algorithm_NAPG_power_control} and $\mathcal{O}(T_{inner,\boldsymbol{\mu}}MK^2)$ is the complexity of finding a feasible step size, where $T_{inner,\boldsymbol{\mu}}$ is the average number of backtracking steps per iteration.

 \subsection{Alternating Optimization} 

In this section, we propose the implementation of the alternating optimization algorithm to iteratively determine both $\textbf{u}_k$ and $\eta_{mk}$, $~\forall m, k$ for downlink sum SE maximization, solving $P_1$ in \eqref{NAPGA_optimization}. The process is summarized in Algorithm \ref{Algorithm_NAPG_All}.
Specifically, we randomly initialize the state
and alternately perform Algorithms  \ref{Algorithm_NAPG_MA} and \ref{Algorithm_NAPG_power_control}
to increase the downlink sum SE. Since $\text{SE}_\text{sum}(\boldsymbol{\mu}, \textbf{u})$ is
non-decreasing over iterations and has an upper bound, Algorithm \ref{Algorithm_NAPG_All} is guaranteed to converge. Meanwhile, the complexity for maximizing the downlink sum SE
depends on the maximum number of iterations, as well as
the complexity required to solve the aforementioned subproblems. Accordingly, The total computational complexity of Algorithm \ref{Algorithm_NAPG_All} is $\mathcal{O}(T_{outer}(T_1(MKNL^2+T_{inner,\textbf{u}}MK^2N)+T_2T_{inner,\boldsymbol{\mu}}MK^2 )$, where $T_{outer}$ is the number of iterations to complete Algorithm \ref{Algorithm_NAPG_All}. Note that the total complexity increases as $M$ and $K$ increase significantly, motivating scalable low-complexity optimization algorithms and user-centric association designs in our future work to maintain the practical feasibility.

\begin{algorithm}[t!]
   \caption{Alternating Optimization to solve $P_1$ in \eqref{NAPGA_optimization}} 
\begin{algorithmic}[1]
\renewcommand{\algorithmicrequire}{\textbf{Inputs:}}
\Require
$\epsilon$ (tolerance), $\boldsymbol{\mu}^0\geq 0$, $\textbf{u}^0\geq 0$, $\alpha>0$, $\delta>0$, $\text{I}_\text{max}$ (maximum
number of iterations), iteration index $i\gets 0$;
\State Initialize $w_{mk}^{0}=\big{|}\big{|}\textbf{d}^H_{mk}(\textbf{u}_k^{0})\big{|}\big{|}^2$, $\eta_{mk}^{0}=(\mu_{mk}^{0})^2/w_{mk}^{0}$, $\forall m,k$.
\Repeat
\State Update $\textbf{u}^{(i+1)}$ by Algorithm \ref{Algorithm_NAPG_MA}.
\State Update $\textbf{v}_{mk}^{(i+1)}=\textbf{d}^H_{mk}(\textbf{u}_k^{(i+1)}),~\forall m, k$.
\State Update $w_{mk}^{(i+1)}=\big{|}\big{|}\textbf{v}_{mk}^{(i+1)}\big{|}\big{|}^2,~\forall m, k$.
\State Update $\boldsymbol{\mu}^{(i+1)}$ by Algorithm \ref{Algorithm_NAPG_power_control}.
\State Update $\boldsymbol{\eta}^{(i+1)}=[\boldsymbol{\eta}_1^{(i+1)};\boldsymbol{\eta}_2^{(i+1)};...;\boldsymbol{\eta}_K^{(i+1)}]\in\mathbb{C}^{MK\times1}$, where
$\eta_{mk}^{(i+1)}=(\mu_{mk}^{(i+1)})^2/w_{mk}^{(i)},~\forall m, k$.
\State Update $\text{SE}_k^{(i+1)},~\forall k$ with $\eta_{mk}^{(i+1)}$ and $\textbf{u}_k^{(i+1)}$, $\forall m,k$.
\State $i\gets i+1$
\Until $i>\text{I}_\text{max}$ or $\big{|} \text{SE}_\text{sum}(\boldsymbol{\eta}^{(i)},\textbf{u}^{(i)})-\text{SE}_\text{sum}(\boldsymbol{\eta}^{(i-1)},\textbf{u}^{(i-1)})\big{|}^2\leq \epsilon$ is satisfied.
\renewcommand{\algorithmicensure}{\textbf{Output:}}
\Ensure
$\boldsymbol{\eta}^\text{opt}=\boldsymbol{\eta}^{(i)}$, $\textbf{u}^\text{opt}=\textbf{u}^{(i)}$.
\end{algorithmic}
\label{Algorithm_NAPG_All}
 \end{algorithm}

\section{Numerical Results}
In this section, we present numerical results illustrating the impact of asynchronous reception on FAS-enabled cell-free massive MIMO systems and introduce performance limits and system design guidelines accordingly. For fairness, we employ the Monte Carlo method to derive the average performance over plenty of channel realizations as $\mathbb{E}_{\{\boldsymbol{\Sigma}_{mk}\}}\{\text{SE}_{k}\},~\forall m, k$. To simplify the notation, in the following figures, we utilize TAS to represent traditional fixed-position antennas in the legends of figures.

\subsection{Parameter Setup}
In this section, a three-slope propagation model is utilized \cite{10032129,7827017}, where APs and users are uniformly and independently distributed within a square region of 200 m $\times$ 200 m.
The response matrix is assumed to be diagonal with independent and identically distributed (i.i.d.) elements following a circularly symmetric complex Gaussian distribution\cite{10243545,11018493}, i.e., $\boldsymbol{\Sigma}_{mk}[1,1]\sim\mathcal{CN}(0,\beta_{mk}\kappa/(\kappa+1))$
and $\boldsymbol{\Sigma}_{mk}[i,i]\sim\mathcal{CN}(0,\beta_{mk}/((\kappa+1)(L-1))),~i=2,3,...,L$\cite{10243545,11018493}. Here, $\kappa$ is the ratio of the average power for Line-of-sight paths to that for non-line-of-sight paths. $\beta_{mk}$ denotes the large-scale fading coefficient between the $m$-th AP and the $k$-th user, and we adopt the path loss model in \cite{7827017} with the relevant settings to introduce the large-scale fading coefficients. The number of transmit and receive paths is the same, i.e., $L_{mk,r}=L_{mk,t}=L,~\forall m,k$\cite{11018493}. 
Unless otherwise stated, FAS positions are located in a square region $\mathcal{C}_k,~\forall k$, where the length of the moving region side is $\lambda$, namely, $d_\text{max}-d_\text{min}=\lambda$ \cite{11018493}. $\kappa=1$ and $L=10$ are assumed for the response matrix, $N_v=N_h$ is adopted for all APs.
The elevation and azimuth AoDs/AoAs are assumed to be i.i.d variables with the uniform distribution over $[-\pi/2,\pi/2]$ \cite{10243545}. The speed of light is $c=3\times10^8$ m/s, the symbol duration is $T_s=1~\mu\text{s} $ for megahertz-bandwidth narrowband services\cite{6862056,9310357}
and the carrier frequency is $f_c=5$ GHz \cite{10103838}. Note that the designated size guarantees that the maximum timing offset is much smaller than the symbol duration, i.e., the maximum timing offset is 0.94 $\mu\text{s}$ for the extreme case when two APs are on the diagonal line of the designated area, and the served user is in the same position as one of the APs.
The downlink transmit power is $p_{d}=23~\text{dBm}$, and the noise power is $\sigma^2=-91$ dBm. The maximum number of iterations is $\text{IterMax}=200$ and the tolerance is $\epsilon=10^{-5}$ for the projected NAPGA algorithm.

\subsection{Convergence Behavior of the NAPGA Algorithm}
\begin{figure}[t!]
    \centering
    \includegraphics[width=\linewidth, height=0.7\linewidth]{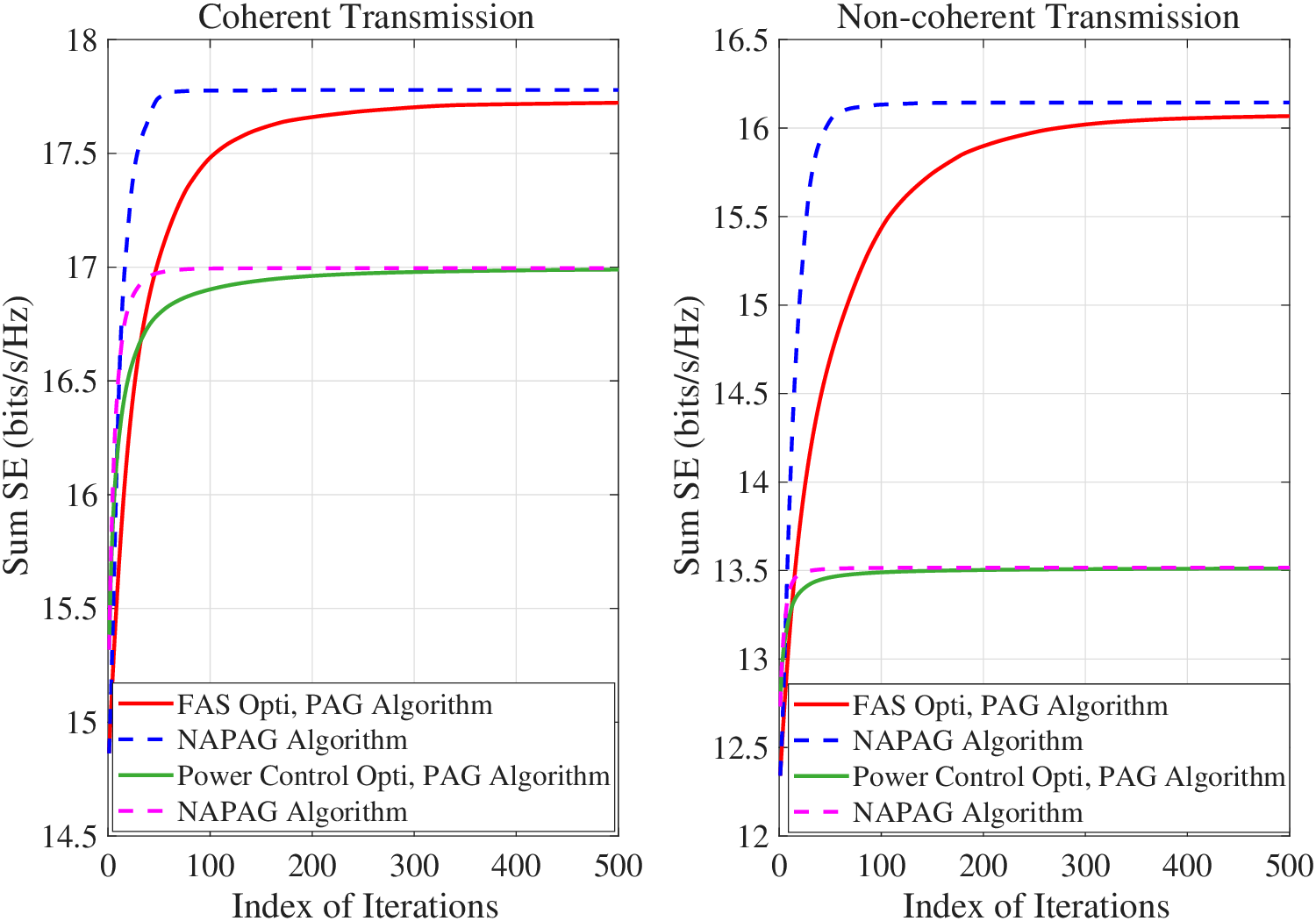}
    \caption{The iterative convergence behaviour of the NAPGA algorithm compared to the conventional PGA algorithm with $M=20$, $N=4$, $K=10$.}
    \label{iteration}
\end{figure}

Fig. \ref{iteration} evaluates the convergence behavior of the proposed
NAPGA algorithm under both coherent and non-coherent transmission schemes, as a function of the iteration index. For comparison, we consider the conventional projected gradient ascent (PGA) algorithm as a non-accelerated baseline \cite{10416363}. The results show that the proposed NAPGA algorithm exhibits a significantly higher convergence rate than the PGA algorithm. Specifically, the NAPGA algorithm converges and achieves near-optimal sum SE within approximately 30 to 50 iterations, whereas the non-accelerated PGA algorithm requires at least 200 iterations to reach a comparable steady state. 
Consequently, the NAPGA curves exhibit a much sharper initial ascent; this convergence gain highlights the efficacy of the acceleration terms in navigating the non-convex optimization landscape inherent in FAS systems and efficient power control.
Moreover, the performance margin between the FAS position optimization and power control optimization curves physically underscores the advantage of spatial reconfigurability introduced by FAS since FAS position optimization physically alters the propagation environment to natively circumvent interference, while power control manages interference over a static, delay phase-degraded channel.
Meanwhile, while coherent transmission yields better performance, FAS position optimization offers greater performance gains for non-coherent transmission since non-coherent transmission can counteract phase asynchronization; thus, the FAS is liberated to reconfigure channels to optimize received signal levels, resulting in substantial relative performance gains compared to fixed-position antennas.
\begin{figure}[t!]
    \centering
    \includegraphics[width=1\linewidth]{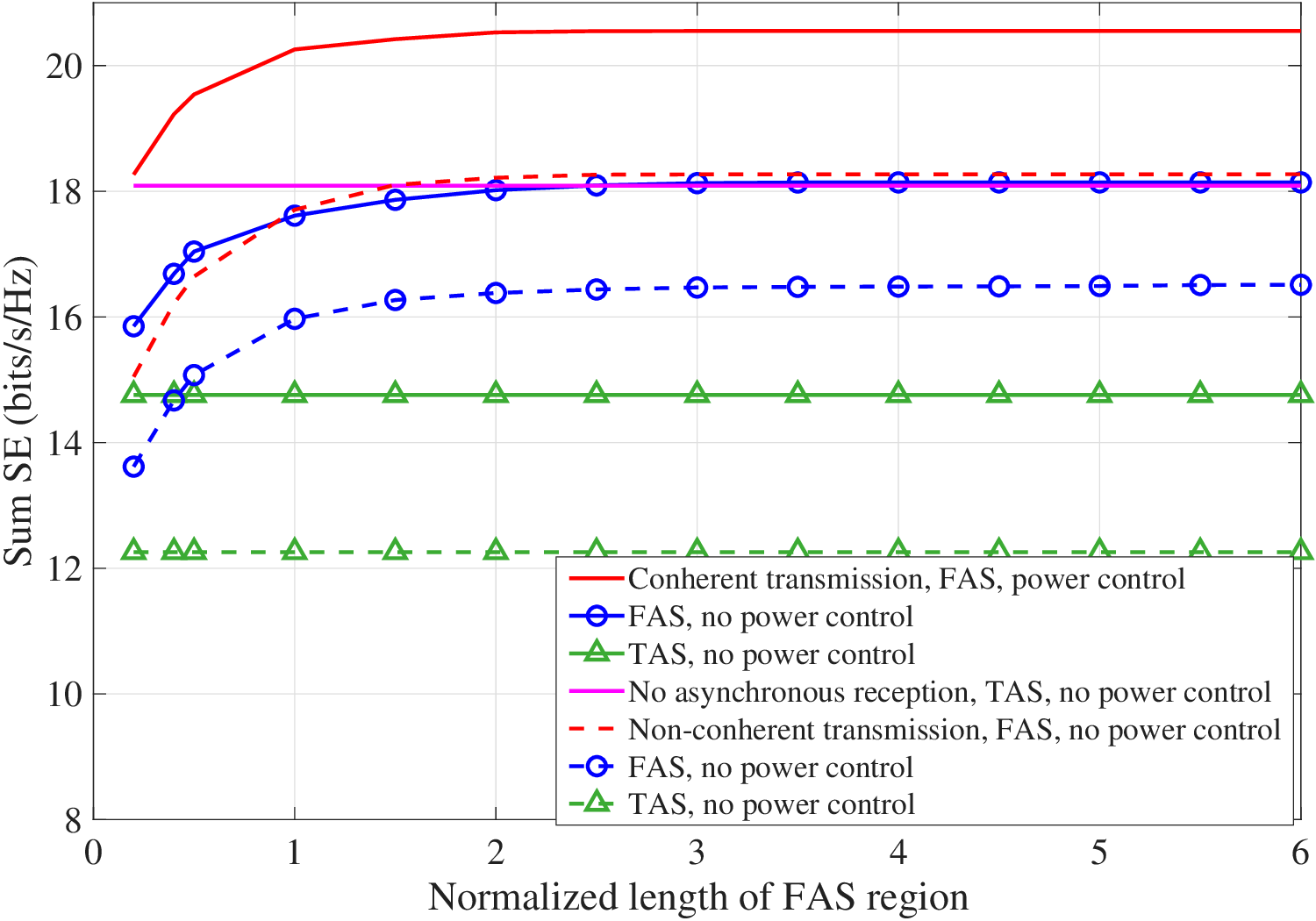}
    \caption{Downlink sum SE vs the normalized length of the FAS region side, $\left(d_\text{max}-d_\text{min}\right)/\lambda$, with $M=20$, $N=4$, $K=10$.}
    \label{FAS}
\end{figure}
\subsection{Effect of FAS Configuration}
Fig. \eqref{FAS} illustrates the downlink sum SE as a function of the designated length of the FAS moving region side, $\left(d_\text{max}-d_\text{min}\right)/\lambda$. 
The results clearly indicate that enlarging the length of the moving region allows the FAS to exploit greater spatial diversity, thereby yielding an enhanced downlink sum SE. For example, compared to $d_\text{max}-d_\text{min}=0.2\lambda$, $d_\text{max}-d_\text{min}=2\lambda$ provides nearly $15\%$ and $20\%$ SE increase for coherent and non-coherent transmissions, respectively. However, the performance gains begin to saturate beyond $d_\text{max}-d_\text{min}>2\lambda$, suggesting that a moderately sized FAS moving region is sufficient to capture most of the available spatial DoFs within the proposed propagation environment. Meanwhile, FAS with $d_\text{max}-d_\text{min}=2\lambda$ can introduce $22\%$ and $34\%$ SE increase for coherent and non-coherent transmissions, respectively. Accordingly, FAS with optimized positions markedly advance the downlink sum SE compared to traditional fixed-position antennas.
Moreover, complementing FAS with power control optimization yields an additional $14\%$ SE increase, proving essential for managing inter-user interference and optimizing the downlink sum SE. We also find that the proposed coherent transmission with joint power control and FAS position optimization can achieve an SE increase of around $10\%$ relative to the baseline performance of perfectly synchronous coherent transmission, and the proposed system with FAS position optimization can approach this baseline performance. This highlights that the immense spatial diversity unlocked by FAS can sufficiently compensate for the destructive delay phases induced by asynchronous reception in coherent transmission, while efficient power control can further improve performance.
Consequently, deploying moderately sized FAS in cell-free massive MIMO systems is a highly robust strategy to achieve substantial improvements. Moreover, the FAS area is determined by the carrier frequency wavelength; thus, a larger moving region can be realized in higher-frequency bands, facilitating better performance\cite{10146262}.

\begin{figure}[t!]
    \centering
    \includegraphics[width=\linewidth]{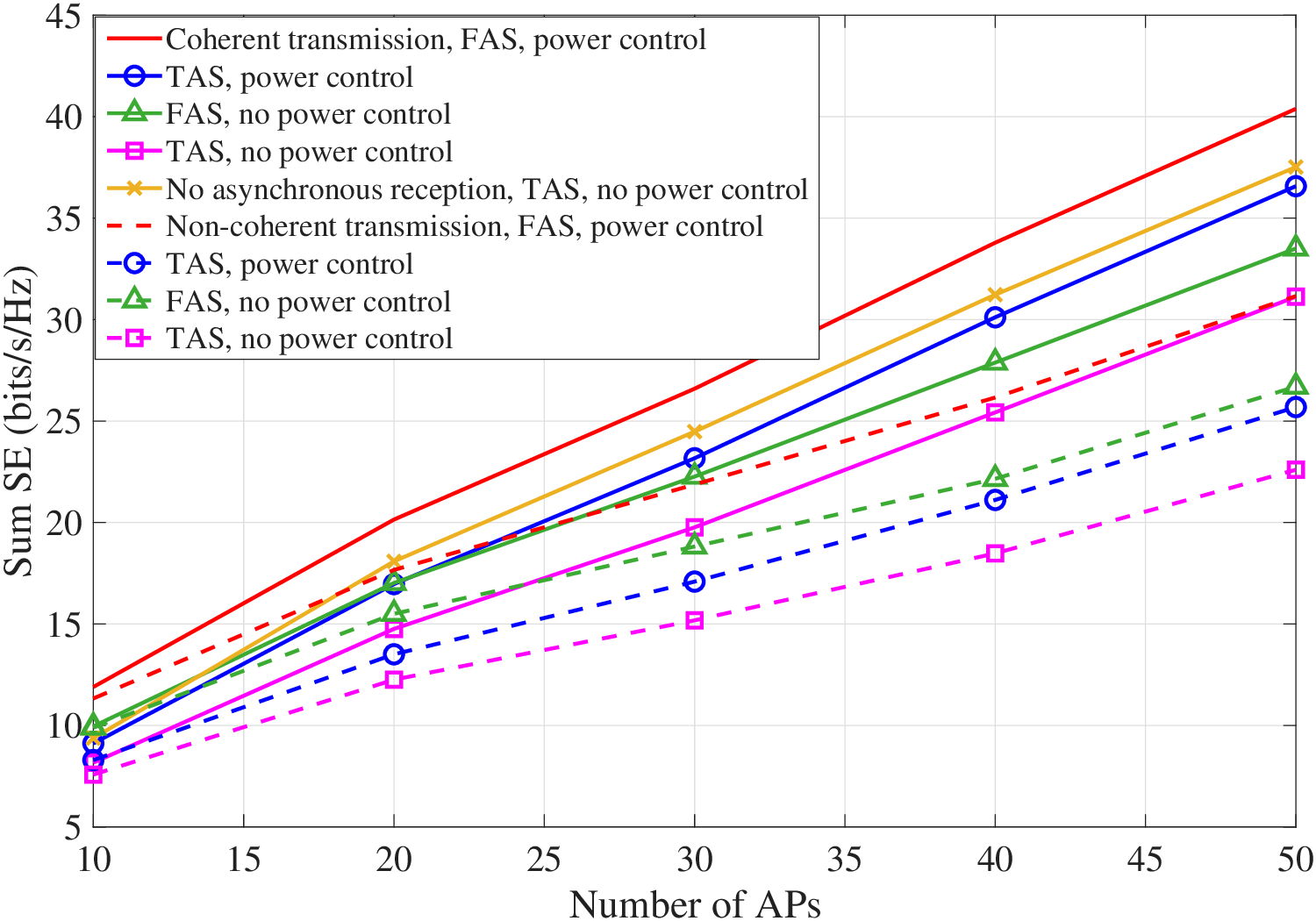}
    \caption{Downlink sum SE vs the number of APs with $N=4$, $K=10$.}
    \label{AP}
    \vspace{-0pt}
\end{figure}
\begin{figure}[t!]
    \centering
    \includegraphics[width=\linewidth]{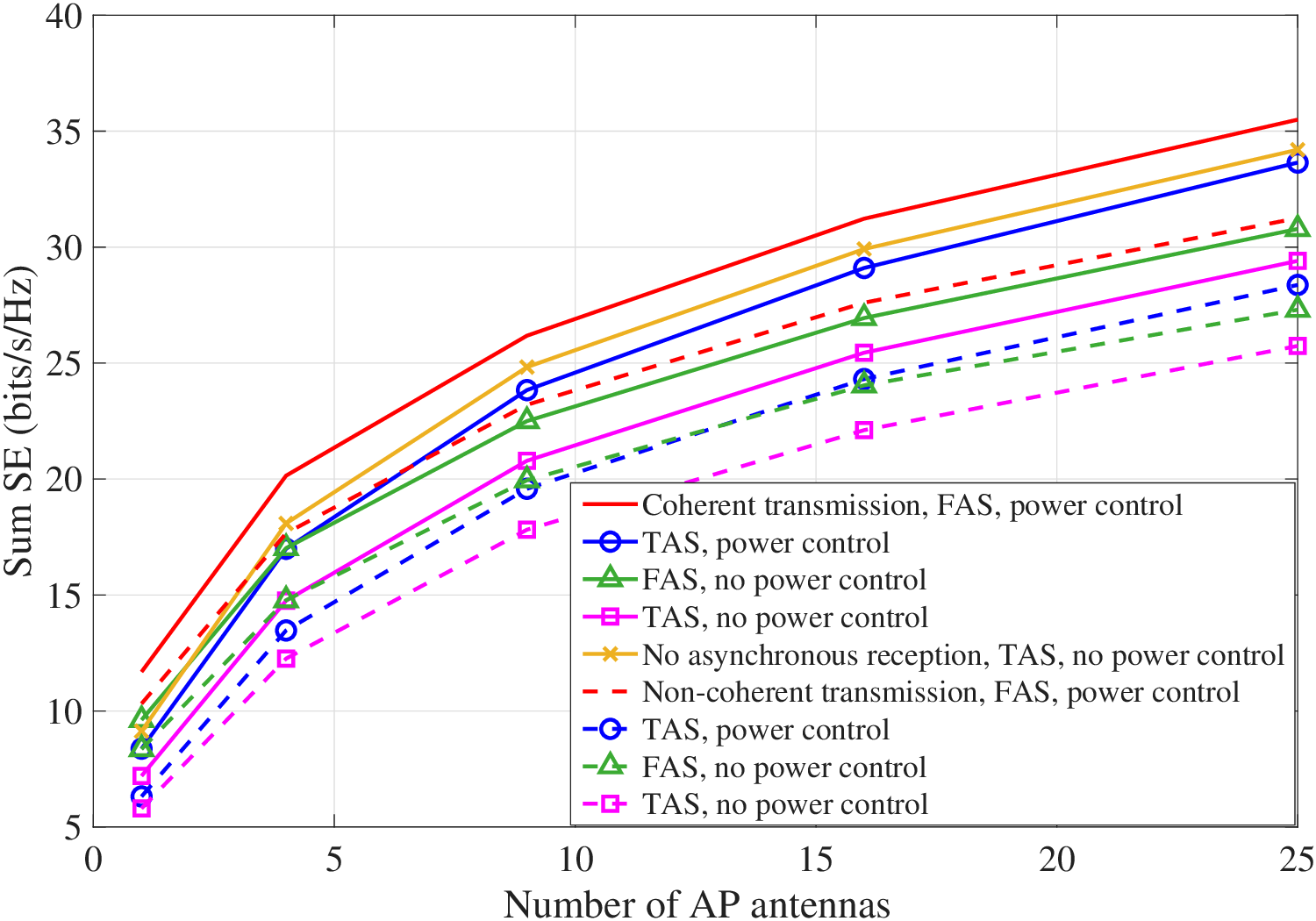}
    \caption{Downlink sum SE vs the number of AP antenna with $M=20$, $K=10$.}
    \label{AP_antenna}
    \vspace{-2pt}
\end{figure}

\subsection{Effect of Number of APs and Number of AP Antennas}
Fig. \ref{AP} depicts the downlink sum SE versus the number of APs, which serves as a fundamental metric for evaluating cell-free massive MIMO systems. 
The results reveal that increasing the number of APs can greatly improve the sum SE, indicating that more APs can significantly increase the spatial DoF, thereby enabling more efficient beamforming. For coherent transmission, power control optimization delivers over $15\%$ SE increase compared to the baseline without power control, and FAS position optimization introduces over $7\%\sim20\%$ SE increase compared to traditional fixed-position antennas. Although the growth rate introduced by FAS decreases as the number of APs increases, the benefits of offsetting asynchronous reception remain remarkable, since increasing the number of APs can yield an SINR increase that is nearly proportional to the number of APs. Meanwhile, power control optimization yields a slightly higher SE gain than FAS position optimization with more APs, indicating that while FAS can reconfigure channels, power control can be more effective at managing the complex inter-user interference across the vast distributed array.
For non-coherent transmission,
power control optimization exhibits over $12\%$ SE increase compared to the baseline without power control, and FAS position optimization delivers over $18\%\sim30\%$ SE increase compared to traditional fixed-position antennas
In contrast to coherent transmission, FAS optimization outperforms power control optimization since non-coherent transmission can counteract asynchronous reception effects and benefit immensely from physically optimizing the FAS to a position that can highly improve received signal levels, proving that spatial reconfigurability dominates the non-coherent transmission performance. Meanwhile, the joint optimization of power control and FAS positions delivers the highest performance, exhibiting a $30\%\sim45\%$ SE increase for coherent transmission and a $37\%\sim50\%$ SE increase for non-coherent transmission compared to the baseline with fixed-position antennas and no power control. Notably, joint optimization under asynchronous reception outperforms perfectly synchronous coherent transmission, demonstrating the necessity of jointly optimizing antenna positions and power control coefficients to mitigate asynchronous reception effects. 

Increasing the number of antennas per AP can greatly improve the downlink SE, as shown in Fig. \ref{AP_antenna}. However, the benefits of deploying FAS with power control diminish as the number of antennas per AP increases for both coherent and non-coherent transmission, following the same trend as in Fig. \ref{AP}. Take coherent transmission for example, when compared with the baseline with fixed-position antennas, FAS position optimization can introduce $33\%$ SE increase when $N_{ap}=1$; however, only $15\%$ SE increase and $8\%$ SE increase can be introduced when $N_{ap}=4$ and $N_{ap}=9$, respectively. Similarly, the joint power control and FAS position optimization can introduce $62\%$ SE increase when $N_{ap}=1$; however, only $36\%$ SE increase and $25\%\sim30\%$ SE increase can be introduced when $N_{ap}=4$ and $N_{ap}=9$, respectively. In contrast, power control optimization maintains a highly consistent relative SE improvement of over $10\%$ regardless of the number of AP antennas. This is because a larger number of antennas per AP enhances transmitter directivity and makes the AP steering vectors mathematically dominate the channel coefficients, thereby weakening the benefits of deploying FAS at users.
In this case, to maximize cost-to-performance efficiency in practical system designs, the number of distributed APs and the number of antennas per AP should be appropriately chosen to fully realize the spatial DoF benefits offered by FAS. 
\subsection{Effect of Number of Users}
\begin{figure}[t!]
    \centering
    \includegraphics[width=\linewidth]{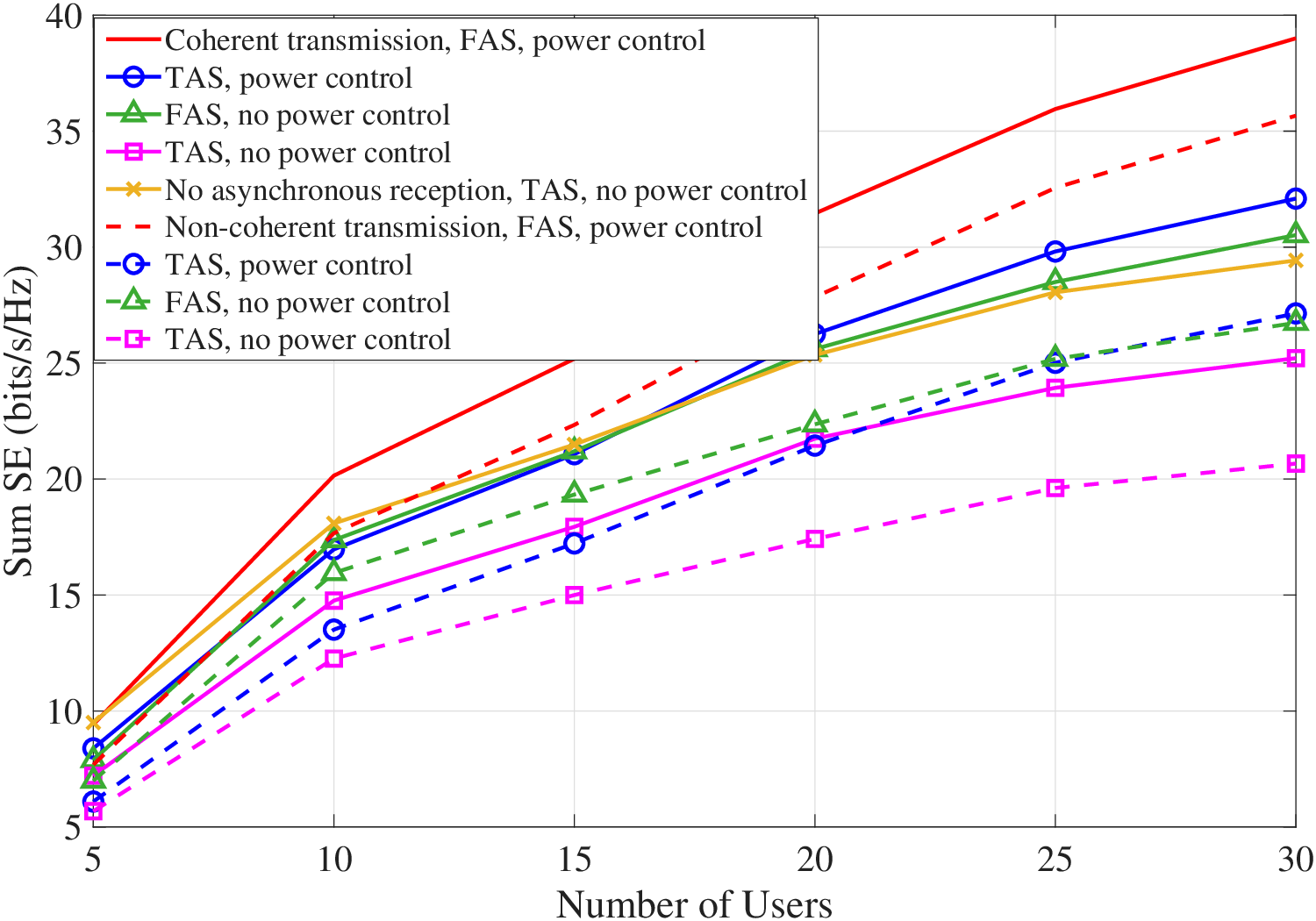}
    \caption{downlink sum SE vs the number of users with $M=20$, $N=4$.}
    \label{user}
\end{figure}
Fig. \eqref{user} displays the downlink sum SE as a function of the number of users. It demonstrates that the downlink sum SE increases monotonically with the increasing number of users by exploiting multi-user spatial multiplexing gains. 
As the number of users increases and the system enters a highly interference-limited regime, efficient power control can significantly boost the downlink sum SE, offsetting asynchronous reception effects and managing inter-user interference, yielding a $16\sim27\%$ SE increase for coherent transmission and $7\sim30\%$ for non-coherent transmission. Similarly, FAS introduce a $9\sim21\%$ SE increase for coherent transmission and $23\sim30\%$ SE increase for non-coherent transmission over traditional fixed-position antennas. 
Ultimately, releasing the synergistic benefits of joint power control and FAS position optimization delivers massive capacity gains, achieving a $30\sim55\%$ SE increase for coherent transmission and $35\sim72\%$ SE increase for non-coherent transmission.
Crucially, as network density increases, this joint optimization under asynchronous reception continues to outperform perfectly synchronized coherent transmission, demonstrating that the proposed joint optimization framework is exceptionally robust to delay phases, especially in dense multi-user scenarios. Despite these benefits, supporting a massive number of users poses practical challenges. As the number of users grows, the joint optimization exerts significant computational complexity, processes overhead and heightens inter-user interference. To maintain scalability and low-complexity feasibility, this FAS-enabled framework should be integrated with an efficient, user-centric architecture alongside other advanced signal processing approaches \cite{10422885}.

\section{Conclusion}

This paper proposed a novel FAS-enabled cell-free massive MIMO system subjected to practical asynchronous reception effects. By empowering users with reconfigurable FASs, we established a mechanism to natively counteract the destructive asynchronous reception arising from inevitable differences in signal arrival times. We developed a comprehensive downlink transmission model and derived the analytical SE expressions for both coherent and non-coherent transmissions, accounting for the impact of delay phases under low-complexity MR precoding. Subsequently, we introduced a novel NAPGA algorithm to jointly optimize the FAS positions and downlink power control coefficients, maximizing the downlink sum SE.
Numerical results verified that while asynchronous reception incurs severe performance degradation, the proposed framework successfully neutralizes these effects.
Specifically, the joint optimization via the NAPGA algorithm effectively compensates for delay phases in coherent transmission, significantly outperforming traditional fixed-position antennas. Conversely, in delay phase-immune non-coherent transmission, FASs are liberated to fully exploit their spatial reconfigurability, yielding even more pronounced SE gains. Our results also revealed that increasing the number of APs and AP antennas induces a localized spatial hardening effect that gradually diminishes the benefits of FAS, while deploying more FAS can effectively release their spatial reconfigurability to improve performance. Thus, a careful system architecture design is essential to maximize system performance. Ultimately, this work demonstrates that the physical adjustment of FAS positions is highly potent for mitigating multi-user interference and asynchronous reception in distributed networks. 

To bridge the gap between the obtained theoretical capacity bounds and practical implementation, our future work will extend the proposed framework to accommodate more realistic and dynamic wireless conditions. 
Specifically, to address practical limitations, we will incorporate channel estimation errors, which can be bounded by spherical uncertainty regions or statistical Gaussian distributions, into the proposed framework. This will involve performance analysis against the perfect CSI baseline and the development of robust, worst-case-bounded optimization designs to enhance system resilience, thereby demonstrating the robustness of the FAS-enabled architecture. Furthermore, we will incorporate the impact of ISI and inter-carrier interference (ICI) induced by significant timing offsets in broadband scenarios, e.g., cell-free Massive MIMO-OFDM systems, by jointly designing advanced ISI/ICI mitigation techniques alongside FAS position optimization.
Finally, integrating scalable user-centric User-AP associations with FAS position optimization will be a crucial extension to maintain practical feasibility and scalability in ultra-dense distributed 6G systems, satisfying the ever-growing demand for wireless data and ubiquitous connectivity.

\ifCLASSOPTIONcaptionsoff
 \newpage
\fi


%




\ifCLASSOPTIONcaptionsoff
 \newpage
\fi



\bibliographystyle{IEEEtran}
\bibliography{IEEEabrv,ref}
\end{document}